\documentclass[sigconf]{acmart}

\AtBeginDocument{%
  }

\setcopyright{acmlicensed}
\copyrightyear{2018}
\acmYear{2018}
\acmDOI{XXXXXXX.XXXXXXX}
\acmConference[Conference acronym 'XX]{Make sure to enter the correct
  conference title from your rights confirmation email}{June 03--05,
  2018}{Woodstock, NY}
\acmISBN{978-1-4503-XXXX-X/2018/06}

\usepackage{algorithm}
\usepackage{algpseudocode}
\usepackage{enumitem}
\usepackage{multirow}
\usepackage{subcaption}
\usepackage[table]{xcolor}

\begin{document}

\title{AnimWeb-Bench: Benchmarking Animated Webpage Code Generation from Videos}



\author{Zhongyi Zhang}
\affiliation{%
  \institution{The Chinese University of Hong Kong}
  \city{Hong Kong}
  \country{China}
}
\email{zyzhang01@link.cuhk.edu.hk}

\author{Jingyu Xiao}
\authornote{Jingyu Xiao is the corresponding author.}
\affiliation{%
  \institution{The Chinese University of Hong Kong}
  \city{Hong Kong}
  \country{China}
}
\email{jyxiao@link.cuhk.edu.hk}

\author{Shuqing Li}
\affiliation{%
  \institution{The Chinese University of Hong Kong}
  \city{Hong Kong}
  \country{China}
}
\email{sqli21@cse.cuhk.edu.hk}

\author{Michael R. Lyu}
\affiliation{%
  \institution{The Chinese University of Hong Kong}
  \city{Hong Kong}
  \country{China}
}
\email{lyu@cse.cuhk.edu.hk}

\renewcommand{\shortauthors}{Zhang et al.}

\begin{abstract}
Animation is pervasive on the Web, the CSS animation property alone appears on over three-quarters of webpages.
Yet existing webpage code generation benchmarks target static screenshots, user-triggered interactions, or isolated motion graphics, leaving the ability of multimodal large language models (MLLMs) and coding agents to reconstruct complete animated webpages largely unexplored. To fill this gap, we introduce AnimWeb-Bench, the first benchmark for animated webpage code generation from videos.
It comprises 124 real-world webpages spanning 7 application domains and 33 animation types, with 91.9\% of pages combining multiple types, each paired with a reference video and the static assets of the original webpage. We further propose a human-aligned, two-dimensional evaluation protocol: Static Appearance Fidelity measures structure, text, color, and visual similarity on aligned keyframes, while Animation Fidelity tracks UI components across frames and compares their position, size, velocity, and appearance with those in the reference video. 
Evaluating 14 MLLMs and three coding agents reveals: (1) reproducing webpage animations is substantially more difficult than reproducing static appearance; (2) explicit frame inspection and render feedback improve the reconstruction quality of coding agents; and (3) rapid, continuous changes are harder to reproduce than discrete state transitions. Our code and dataset are available at \href{https://starry1001.github.io/AnimWeb-Bench-website/}{starry1001.github.io/AnimWeb-Bench-website}.
\end{abstract}


  

\received{20 February 2007}
\received[revised]{12 March 2009}
\received[accepted]{5 June 2009}

\maketitle

\section{Introduction}

Motion is part of the Web's basic visual vocabulary. The CSS \texttt{animation} and \texttt{transition} properties appear on 77\% and 85\% of the nearly 8 million mobile pages crawled by the HTTP Archive~\cite{webalmanac2022css}, and about half of all pages ship a \texttt{prefers-reduced-motion} query to adapt that motion to users' needs~\cite{webalmanac2025a11y}. 
On these pages, animation is functional rather than decorative: it signals system feedback, preserves continuity across state changes, and directs attention~\cite{chang1993animation,heer2007animated}. 
At the same time, front-end code is increasingly written by Multimodal Large Language Models and coding agents: 84\% of developers use or plan to use AI tools in their development process~\cite{stackoverflow2025survey,zhao2026beyond}, and generating a webpage from a visual reference has become a prominent use case targeted by a growing line of benchmarks~\cite{si2025design2code,gui2025webcode2m}.

Several benchmarks have been proposed to evaluate MLLMs on front-end code generation, but most of them focus on static webpage reconstruction from screenshots~\cite{laurencon2024unlocking,yun2024web2code,si2025design2code,gui2025webcode2m}. A few studies extend the evaluation to user-triggered interactions. Interaction2Code~\cite{xiao2024interaction2code} represents each interaction with screenshots captured before and after a user action, while IWR-Bench~\cite{chen2026iwr} and WebVR~\cite{dai2026webvrbenchmarkingmultimodalllms} use demonstration videos to evaluate the reconstruction of interactive webpages. However, these benchmarks primarily target discrete state changes triggered by user actions rather than animations that unfold continuously over time.

This raises a question that existing evaluations leave open: \emph{can AI systems reconstruct not only the visual states of a webpage, but also the continuous dynamics connecting them?}

\begin{figure*}[htbp]
  \centering
  \includegraphics[width=0.9\textwidth]{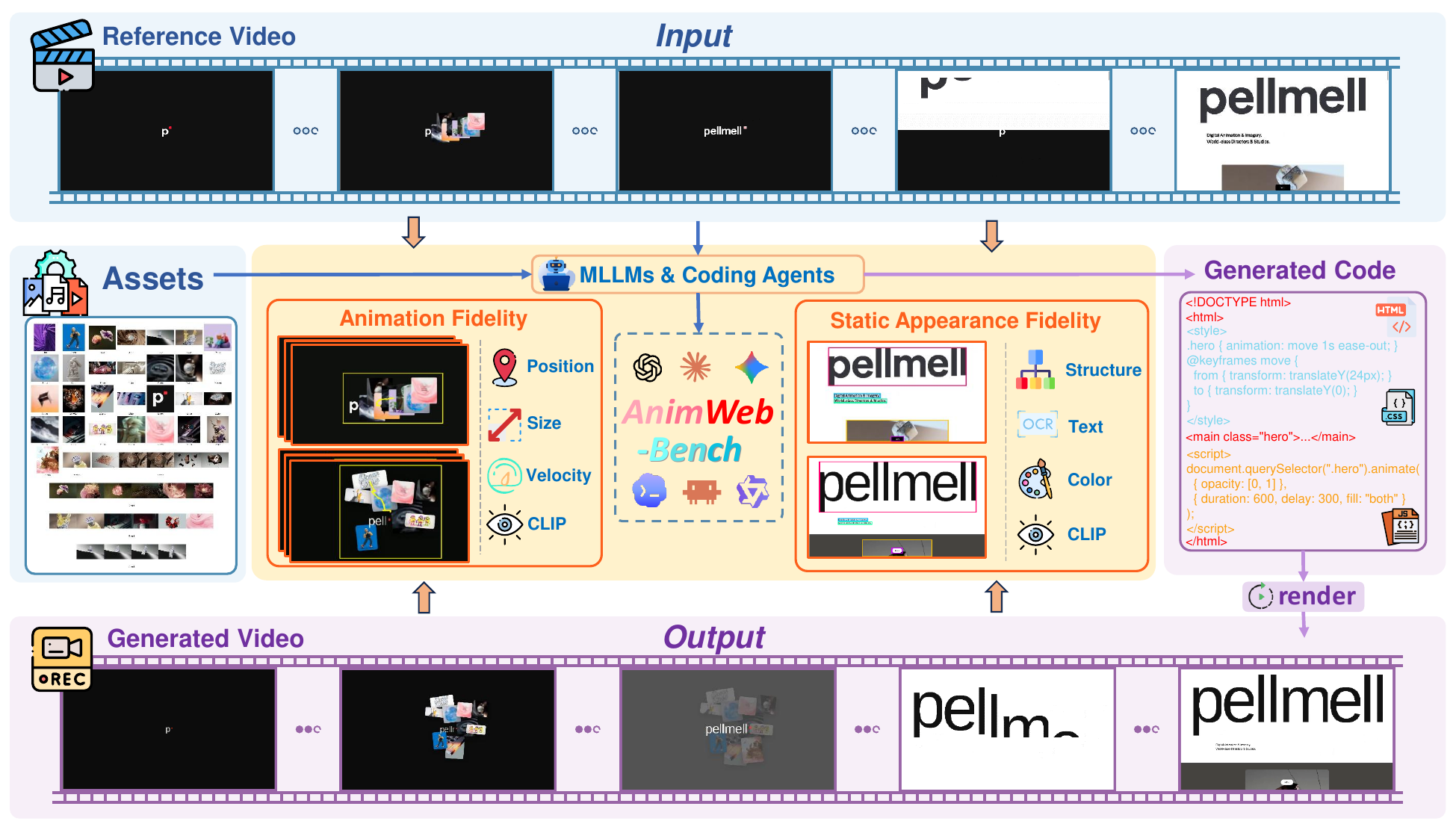}
  \caption{Overview of the AnimWeb-Bench task and evaluation. Given a reference video and the static assets of a real webpage (top), an MLLM or coding agent generates a single HTML file, which is rendered and recorded in a browser (bottom). Static Appearance Fidelity compares aligned keyframes. Animation Fidelity compares the trajectories of matched UI components.}
  \label{fig:overview}
\end{figure*}

Answering this question requires moving beyond static appearance and discrete interactive functionality to evaluate the continuous temporal behavior of visual elements. This poses additional challenges: a model must infer motion trajectories, timing, easing, and synchronization from the reference video and express them in executable code. Animation2Code~\cite{ji2026animation2codeevaluatingtemporalvisual} took a first step answering it, reconstructing web animation code from videos, but on isolated motion graphics scored within a cropped region.

Real webpages are harder. Many components move in sequence with distinct timing and easing, while coexisting with the page's layout, text, and media.
Reconstruction requires inferring which components move, when, and how fast. Evaluation requires judging the motion of each component, which frame-level similarity cannot see, region-pooled tracks cannot attribute, and binary rubrics cannot grade.

To bridge this gap, we introduce AnimWeb-Bench, to our knowledge the first benchmark for reconstructing complete real-world webpages together with their time-driven animations from videos. 
AnimWeb-Bench comprises 124 real-world webpages spanning 7 application domains and 33 animation types. 
As a comparison, an AnimWeb-Bench page contains 1,196 HTML tags on average, versus 71 for an Animation2Code sample. 
As illustrated in Figure~\ref{fig:overview}, each task pairs a reference video with the original webpage's static assets.
To evaluate the generated webpages, we design a human-aligned protocol with two complementary dimensions: \emph{Static Appearance Fidelity}, which compares structure, text, color, and visual similarity on aligned keyframes, and \emph{Animation Fidelity}, which tracks UI elements across frames and compares their position, size, velocity, and appearance with the reference.
All eight metrics correlate with 400 blind pairwise human judgments (Spearman's $\rho = 0.67$--$0.74$).

We evaluate 14 MLLMs and 3 coding agents on AnimWeb-Bench. Beyond overall performance, we investigate whether coding agents improve over their underlying models and what factors affect their performance, and analyze how different animation types affect reconstruction. Our main contributions are as follows:
\begin{itemize}[leftmargin=*]
\item 
We introduce AnimWeb-Bench, a benchmark for animated webpage code generation from videos. It comprises 124 real-world webpages spanning 7 application domains and 33 animation types, each paired with a reference video and the original webpage's static assets.
\item 
We design a human-aligned evaluation protocol that separately measures how a generated page looks and how it moves.
\item 
Our comprehensive evaluation of 14 MLLMs and 3 coding agents reveals three key findings: (1) reproducing webpage animations remains challenging for current systems; (2) explicit frame inspection and render feedback improve the reconstruction quality of coding agents; (3) rapid, continuous changes are harder to reproduce than discrete state transitions.
\end{itemize}

\section{Background}

\subsection{Task Definition}

The Animated Webpage Code Generation task aims to reconstruct executable
webpage code by observing the animation effects shown in a reference
video. Given a reference video $V^r = \{I_1, I_2, \ldots, I_n\}$
and a set of static assets (e.g., icons, images and embedded videos): $A = \{a_1, a_2, \ldots, a_m\}$, a model $M$
generates a single executable HTML file $C^g = M(V^r, A)$, with CSS and
JavaScript embedded inline. Rendering and recording $C^g$ in a browser
produces a video $V^g = R(C^g, A)$ (Figure~\ref{fig:overview}). The generated video $V^g$ should
closely approximate $V^r$ in both visual appearance and animation
dynamics.

\subsection{Related Work}

\textbf{UI Code Generation Methods.}
Recent advances in multimodal large language models (MLLMs) have spurred a range of methods for generating webpage code from visual inputs.
DCGen~\citep{wan2024automatically} adopts a divide-and-conquer strategy, DeclarUI~\citep{zhou2024bridging} integrates segmentation with page transition graphs, and UICopilot~\citep{gui2025UICopilot}, LayoutCoder~\citep{wu2024mllm}, and LaTCoder~\citep{gui2025latcoder} enhance generation through hierarchical or layout-aware modeling.
Instruct4Edit~\citep{dang2025envisioning} provides
a scalable and transparent foundation for natural language-based web editing. ComUICoder~\citep{xiao2026component} promotes component reuse, while EfficientUICoder~\citep{xiao2026efficientuicoder} reduces redundant visual and code tokens.
More recently, Video2Code~\citep{xu2026video2code} and InterWebCoder~\citep{xiao2026interwebcoder} extend generation to interactive webpages, focusing on functional correctness and action-triggered state transitions. TTDev~\citep{wan2026TDDev} applies test-driven development method for full-stack website generation and VisualRepair~\citep{xiao2026visualrepair} proposes dynamic tool calling and region focusing for front-end bug repair.
Our work complements these efforts by evaluating the reconstruction of continuous webpage animations from videos.

\textbf{UI Code Generation Benchmarks.}
Early benchmarks focus on static webpage reconstruction from screenshots.
WebSight~\citep{laurencon2024unlocking} and Web2Code~\citep{yun2024web2code} study large-scale HTML generation using synthetic webpages, while Design2Code~\citep{si2025design2code} and WebCode2M~\citep{gui2025webcode2m} introduce real-world data at increasing scales.
MRWeb~\citep{wan2024mrweb} extends evaluation to multi-page websites with external resources, and DesignBench~\citep{xiao2025designbench} covers generation, editing, and repair across multiple web frameworks. UIBenchkit~\citep{le2026uibenchkit} provides a unified toolkit for UI2Code methods evaluation.

However, screenshot-based benchmarks cannot evaluate temporal behavior.
Interaction2Code~\citep{xiao2024interaction2code} represents individual interactions with before-and-after screenshots, whereas IWR-Bench~\citep{chen2026iwr} and WebVR~\citep{dai2026webvrbenchmarkingmultimodalllms} use demonstration videos to evaluate event-driven interactions and stateful workflows.
These benchmarks primarily capture discrete transitions triggered by user actions, while web animations require recovering continuous dynamics such as trajectories, timing, speed, and easing.
Animation2Code~\citep{ji2026animation2codeevaluatingtemporalvisual} directly studies video-to-animation-code generation but focuses on isolated motion graphics rather than complete webpages.
To our knowledge, AnimWeb-Bench is the first benchmark for reconstructing complete animated webpages from videos.
Table~\ref{tb_webpage_reconstruction} summarizes its differences from existing benchmarks.

\begin{table}[t]
\caption{Overview of existing UI-to-code benchmarks.}
\label{tb_webpage_reconstruction}
\centering
\resizebox{\linewidth}{!}{
\setlength{\tabcolsep}{1.5pt}
\begin{tabular}{lllrl}
\toprule
Benchmark & Input & Data Source & Samples & Target Domain \\
\midrule
Pix2Code~\cite{beltramelli2018pix2code}          & Image       & Synthesized   & 1742  & Static Layouts \\
Web2Code~\cite{yun2024web2code}           & Image       & Synthesized   & 60K   & Static Layouts \\
WebSight~\cite{laurencon2024unlocking}           & Image       & Synthesized   & 2M    & Static Layouts \\
Design2Code~\cite{si2025design2code}             & Image       & C4            & 484   & Static Layouts \\
WebCode2M~\cite{gui2025webcode2m}                & Image       & Common Crawl  & 2.56M & Static Layouts \\
CC-HARD~\cite{gui2025latcoder}                   & Image       & Common Crawl  & 128   & Static Layouts \\
ScreenBench~\cite{jiang2025screencoder}          & Image       & Live Websites & 1,000 & Static Layouts \\
Interaction2Code~\cite{xiao2024interaction2code} & Image Pairs & C4 \& GitHub  & 127   & Interactive Webpages \\
WebVR~\cite{dai2026webvrbenchmarkingmultimodalllms} & Video    & Synthesized   & 175   & Interactive Webpages \\
IWR-Bench~\cite{chen2026iwr}                     & Video       & Live Websites & 113   & Interactive Webpages \\
Animation2Code~\cite{ji2026animation2codeevaluatingtemporalvisual} & Video & CodePen & 1,069 & Motion Graphics \\
\midrule
\textbf{AnimWeb-Bench (Ours)}                    & Video       & Live Websites & 124   & Animated Webpages \\
\bottomrule
\end{tabular}}
\end{table}

\section{AnimWeb-Bench}

\begin{figure}[t!]
    \centering
    \includegraphics[width=\columnwidth]{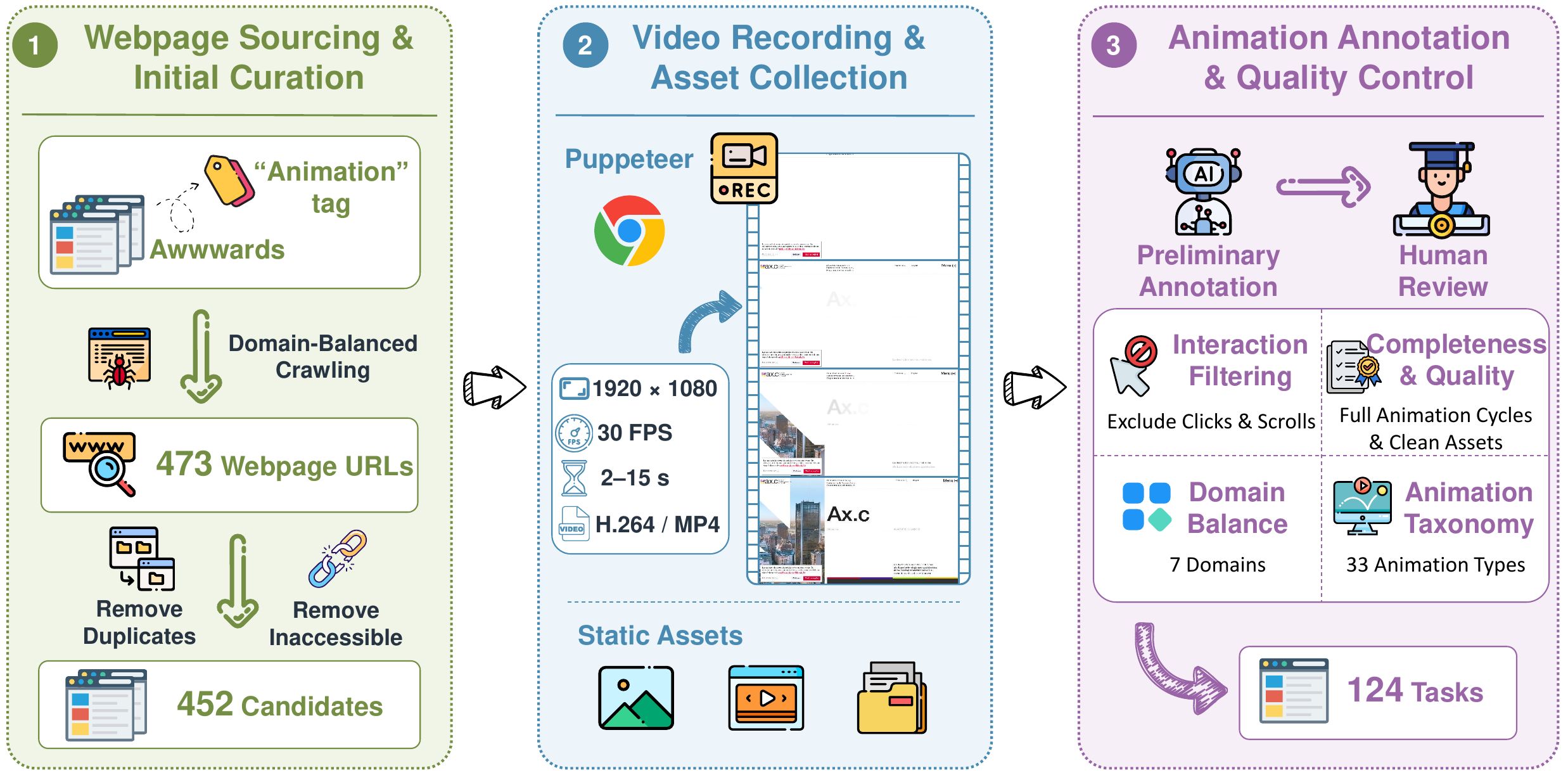}
    \caption{Construction pipeline of AnimWeb-Bench.}
    \label{fig:dataset-construction}
\end{figure}

\subsection{Dataset Construction}
Figure~\ref{fig:dataset-construction} shows the construction pipeline of AnimWeb-Bench.

\textbf{Scope.}
AnimWeb-Bench focuses on \emph{time-driven, page-embedded} animations that unfold automatically during the initial page load. We consider animations implemented through DOM elements using CSS or JavaScript, where motion is integrated with the webpage's layout, text, and media. 


\textbf{Webpage sourcing.} We draw webpages from Awwwards~\cite{awwwards}, a web-design showcase whose entries are reviewed by a jury of professional designers and developers. Starting from sites carrying the \textit{Animation} tag, we crawled up to 100 sites per application domain, collecting 473 URLs of sites submitted between 2025 and 2026. Removing duplicate and inaccessible sites left 452 candidates.

\textbf{Recording and asset collection.} We load each page in headless Chromium (v151.0.7922.34) driven by Puppeteer~\cite{puppeteer} with a $1920\times1080$ viewport and record it at 30\,fps (H.264, MP4) from navigation start. Each recording lasts 2--15\,s (mean 5.91\,s) and covers the page's initial animation sequence, or one full cycle of a looping animation.
We also crawl and deduplicate the static assets needed for reconstruction, i.e., images, icons, and embedded videos (46.1 per page on average).

\textbf{Annotation and quality control.} We first use GPT-5.6 Sol to propose a domain and a set of animation-type labels for each recording from the video and the page source. Three PhD students with front-end development experience then review every instance: they correct the labels, supplement missing assets, and verify that the recording captures the complete animation. 

\subsection{Data Statistics and Diversity}

\begin{figure}[t!]
    \centering
    \includegraphics[width=0.38\textwidth]{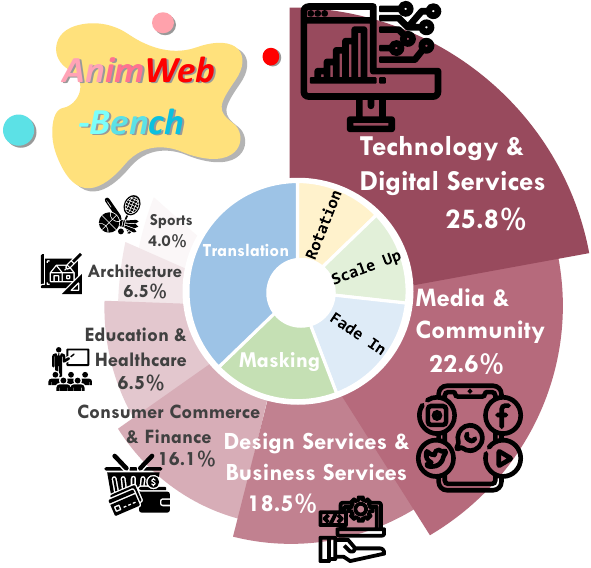} 
    \caption{Domain distribution of AnimWeb-Bench.}
    \label{fig:Data_Diversity}
\end{figure}

\textbf{Domain Diversity.}
As shown in Figure~\ref{fig:Data_Diversity}, AnimWeb-Bench covers 7 application domains, 
the largest groups are technology and digital services (25.8\%), media and community (22.6\%), and design and business services (18.5\%).

\begin{table*}[t!]
\caption{Distribution of animation types in AnimWeb-Bench.}
\label{tb_animation_type_distribution}
\centering
\small
\setlength{\tabcolsep}{3pt}
\resizebox{0.9\linewidth}{!}{%
\renewcommand{\arraystretch}{0.9}
\begin{tabular}{
    @{}lr
    @{\hspace{12pt}}lr
    @{\hspace{12pt}}lr
    @{\hspace{12pt}}lr@{}
}
\toprule
Type & Number &
Type & Number &
Type & Number &
Type & Number \\
\midrule
Translation           & 64 &
Fade Out              & 12 &
Blur Transition       & 4 &
Font Weight Animation & 2 \\

Background Transition & 50 &
Rotation              & 11 &
Icon State Animation  & 4 &
Horizontal Image Slide & 2 \\

Appearance            & 49 &
Numeric Counter Animation & 10 &
Shake                  & 4 &
Vertical Image Slide   & 2 \\

Masking               & 32 &
Scale Down            & 8 &
3D Depth Motion       & 3 &
Logo Morphing         & 1 \\

Fade In               & 30 &
Text Substitution     & 7 &
Grid Tile Transition  & 3 &
Radial Expansion Animation & 1 \\

Text Animation        & 24 &
Image Stack Animation & 5 &
Icon State Transition & 3 &
Text Fragmentation    & 1 \\

Scale Up              & 22 &
Typewriter Animation  & 5 &
Text Scramble         & 3 &
                      &   \\

Color Transition      & 14 &
Vertical Stripe Reveal & 5 &
Twinkling             & 3 &
                      &   \\

Line Growth           & 13 &
Appearance \& Opacity Changes & 5 &
Card Flip             & 2 &
                      &   \\
\bottomrule
\end{tabular}%
}
\end{table*}

\textbf{Animation Type Diversity.}
Based on the preliminary annotations and subsequent manual review, we identify 33 animation types, as detailed in Table~\ref{tb_animation_type_distribution}. 
The most frequent are translation (64), background transition (50), appearance (49), masking (32), and fade-in (30).
In total, the benchmark contains 404 animation-type annotations, 3.3 per webpage on average 
and between one and eight per webpage. 
In particular, 91.9\% of webpages contain more than one animation type, reflecting the complexity of combining multiple effects within a single webpage.

\subsection{Evaluation Metrics}
\label{sec:evaluation}

Our evaluation protocol compares the generated video $V^g$ with the reference video $V^r$ in two complementary dimensions: static appearance and animation.
We first identify keyframes and animation segments in the reference video and align them with the generated video (Section~\ref{sec:keyframes}).
We then evaluate static appearance on pairs of matched keyframes (Section~\ref{sec:static}) and animation on the segments between consecutive matched keyframes (Section~\ref{sec:animation}).

\subsubsection{Keyframes and Animation Segments}
\label{sec:keyframes}

A webpage animation is a continuous process, but treating every video frame independently leads to redundancy and obscures the underlying state transitions.
To obtain a compact and structured representation, we decompose each video into \emph{stable states} and the \emph{animation phases} that connect them. 
This decomposition separates the visual appearance of each state from the dynamic behavior that transforms one state into the next, which our two evaluation dimensions assess separately.

\begin{figure}[t!]
  \centering
  \includegraphics[width=\columnwidth]{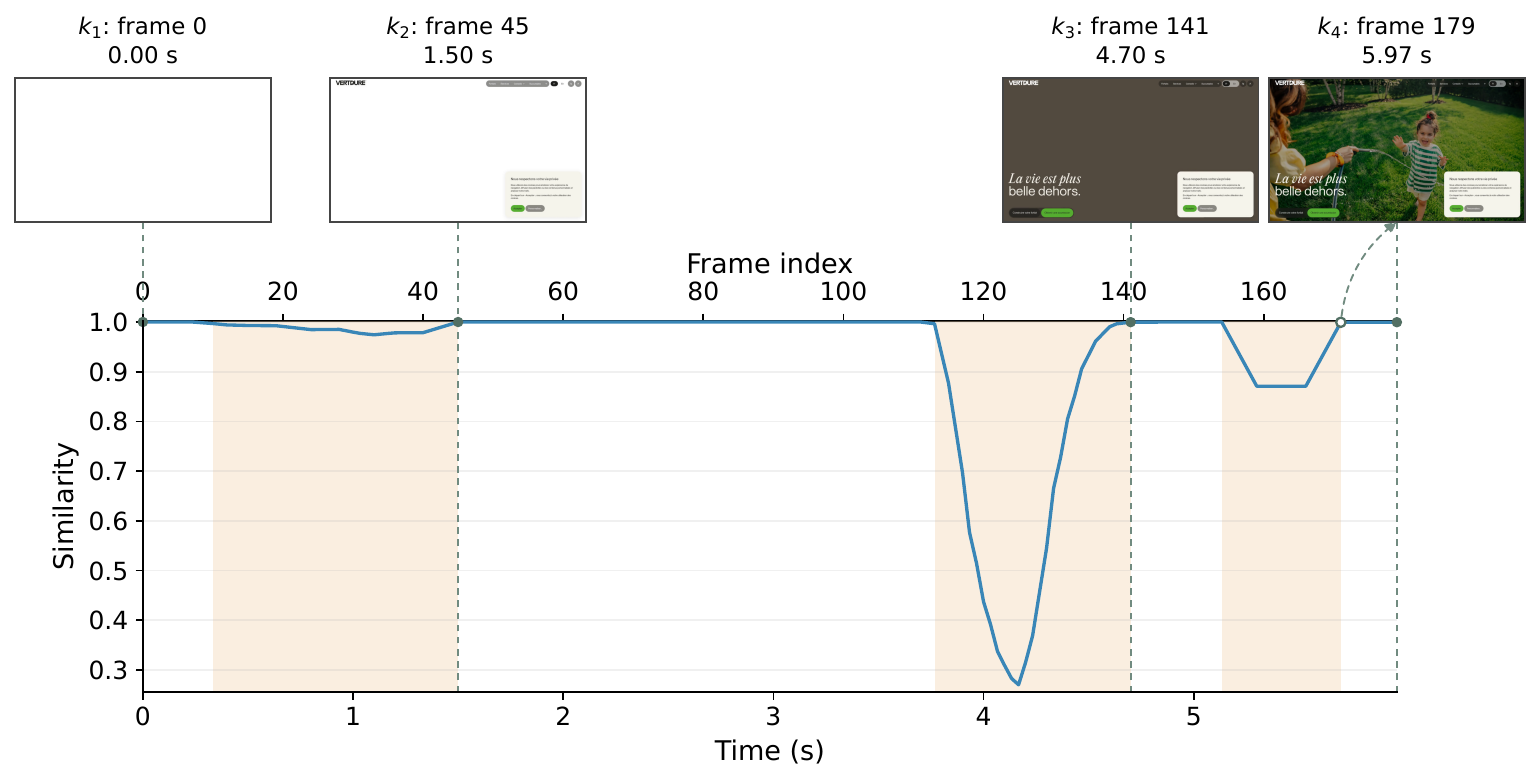}
  \caption{Keyframe extraction on an example from AnimWeb-Bench. The last two detected keyframes have nearly identical content and are therefore merged into $k_4$.}
  \label{fig:keyframe}
\end{figure}

We define a \emph{keyframe} as a frame where the page has settled into a stable state, and an \emph{animation segment} as the interval between two consecutive keyframes that contains the animation leading from one state to the next. To locate keyframes, we measure the degree of visual change around each frame.
For each frame $I_t$ of a video, we compute the similarity between two frames separated by a short temporal window:
\begin{equation}
S(t) = 1 - \frac{1}{|\Omega|}\sum_{p \in \Omega}\frac{\lvert I_{t+h}(p) - I_{t-h}(p) \rvert}{255},
\end{equation}
where $I_t(p) \in [0, 255]$ denotes the grayscale intensity of pixel $p$ in frame $t$, $\Omega$ is the set of pixels of the downsampled frames, and $h$ is the number of frames spanning 0.2 seconds.
The second term is the mean absolute intensity difference between the two frames, normalized to $[0, 1]$, so $S(t)$ approaches 1 when the page is unchanged and decreases as the change grows.
Comparing frames across a window rather than adjacent frames ensures that slow animations, whose per-frame changes are small, remain visible in the similarity curve.
As illustrated in Figure~\ref{fig:keyframe}, $S(t)$ stays close to 1 while the page is stable and drops while an animation is in progress, so each valley of the curve corresponds to an animation phase.
Between two consecutive valleys, the curve returns to a plateau, and we take the first frame on this plateau, i.e., the moment at which the page settles, as a keyframe.
Together with the first and last frames of the video, which capture the initial and final states, these frames form the keyframe sequence $K = \{k_1, \ldots, k_N\}$, in which adjacent keyframes with nearly identical content are merged.


For each keyframe $I^r_k$ extracted from the reference video $V^r$, we locate its counterpart in the generated video $V^g$ by scoring every generated frame $I^g_t$ with
\begin{equation}
\label{eq:kf}
s(I^r_k, I^g_t) = \lambda\,\mathrm{Vis}(I^r_k, I^g_t) + (1-\lambda)\bigl(1 - |\hat{t}^r_k - \hat{t}^g_t|\bigr),
\end{equation}
where $\mathrm{Vis}$ is a low-level visual similarity, $\hat{t}$ is the timestamp normalized by the video duration, and $\lambda$ balances appearance against temporal position. An order-preserving alignment then selects at most one generated frame for each reference keyframe (Appendix~\ref{app:implementation}).
This yields matched keyframe pairs and the corresponding animation segments between consecutive pairs.

\subsubsection{Static Appearance Fidelity}
\label{sec:static}
We compare aligned reference and generated keyframes using
Structure, Text, and Color Similarity as fine-grained metrics, and
Visual Similarity as a high-level metric. Each metric is averaged over
all reference keyframes.

\textbf{Structure Similarity.}
We detect UI components in the reference and generated keyframes with
GPA-GUI-Detector~\citep{salesforce2026gpa}, obtaining box sets $B^r$ and
$B^g$, and match them one-to-one with the element matching algorithm of
ComUICoder~\citep{xiao2026component}, yielding the matched pairs
$\mathcal{M}$. For each pair $(b^r, b^g) \in \mathcal{M}$, we compute
\begin{equation}
\label{eq:structure-pair}
s(b^r,b^g) = 0.35\,s_{\mathrm{iou}} + 0.30\,s_{\mathrm{center}}
+ 0.25\,s_{\mathrm{area}} + 0.10\,s_{\mathrm{aspect}},
\end{equation}
where the four terms measure box overlap, center proximity, area ratio,
and the ratio of aspect ratios on frame-normalized boxes (Appendix~\ref{app:implementation}).
The structure
similarity is
\begin{equation}
S_{\mathrm{str}} = \frac{\sum_{(b^r,b^g)\in\mathcal{M}} s(b^r,b^g)}
{|B^r| + |B^g| - |\mathcal{M}|},
\end{equation}
where the denominator counts each matched pair once, so missing or
redundant components lower the score.


\textbf{Text Similarity.}
We recognize the text in each keyframe with EasyOCR~\citep{easyocr}, concatenate it in reading order, and compute the normalized Levenshtein similarity~\citep{levenshtein1966binary} between the two strings.

\textbf{Color Similarity.}
We treat the components detected in the structure step as foreground and
the remaining regions as background. Following prior
work~\citep{si2025design2code,wan2024mrweb}, we use the CIEDE2000
formula~\citep{luo2001development} to measure the perceptual color
difference of the background and of each matched foreground component.
The final score averages the background and foreground results.

\textbf{Visual Similarity.}
We use the CLIP score~\citep{radford2021learning} to measure the visual
similarity between the reference and generated keyframes.

\subsubsection{Animation Fidelity}
\label{sec:animation}

\begin{algorithm}[t]
\caption{Trajectory Construction and Matching}
\label{alg:trajectory}
\begin{algorithmic}[1]
\Require Reference segment $\mathcal{V}^r$
\Require Generated segment $\mathcal{V}^g$
\Require GPA-GUI-Detector $\mathcal{D}$
\Require Thresholds $\theta_a$, $\theta_t$
\Ensure Trajectory sets $\mathcal{T}^r$, $\mathcal{T}^g$
\Ensure Matched trajectory pairs $\mathcal{A}$
\Statex \textit{Stage 1: Within-video trajectory construction}
\For{$m \in \{r, g\}$}
  \State $\mathcal{T}^m \gets \emptyset$
  \For{each frame $I_t \in \mathcal{V}^m$}
    \State $B_t \gets \mathcal{D}(I_t)$
    \For{each $\tau \in \mathcal{T}^m$ and $b \in B_t$}
      \State $a_{\tau b} \gets \alpha \cdot \mathrm{IoU}
             + \beta \cdot \mathrm{App} + \gamma \cdot \mathrm{Ctr}$
    \EndFor
    \State Match $\mathcal{T}^m$ and $B_t$ one to one with $a_{\tau b} > \theta_a$
    \State Extend matched $\tau$; add unmatched $b$ to $\mathcal{T}^m$
  \EndFor
\EndFor
\Statex \textit{Stage 2: Cross-video trajectory matching}
\State $\mathcal{A} \gets$ pairs with $\mathrm{Text}(\tau^r_i, \tau^g_j) > \theta_t$
\For{each unmatched $\tau^r_i \in \mathcal{T}^r$ and $\tau^g_j \in \mathcal{T}^g$}
  \State $d_{ij} \gets w_p (1 - P_{ij}) + w_s (1 - S_{ij}) + w_v (1 - V_{ij}) + w_c (1 - C_{ij})$
\EndFor
\State $\mathcal{A} \gets \mathcal{A} \cup \mathrm{Hungarian}(\{d_{ij}\})$
\State \Return $\mathcal{T}^r$, $\mathcal{T}^g$, $\mathcal{A}$
\end{algorithmic}
\end{algorithm}

We compare aligned reference and generated segments, each spanning two consecutive keyframes, using Position, Size, and Velocity Similarity as geometric motion metrics and the CLIP score as an appearance metric. Each metric is averaged over all reference segments.

\textbf{Component Trajectories.}
We apply GPA-GUI-Detector~\citep{salesforce2026gpa} to every frame of a
segment and link detections across adjacent frames into component
trajectories (Stage~1 of Algorithm~\ref{alg:trajectory}). A detection $b$
is linked to a trajectory $\tau$ according to an association score
$a_{\tau b}$, which combines the IoU, appearance similarity
($\mathrm{App}$), and center proximity ($\mathrm{Ctr}$) between $b$ and
the latest box of $\tau$, with weights $\alpha=0.5$, $\beta=0.3$, and
$\gamma=0.2$ (Appendix~\ref{app:implementation}).
For a component $i$, let $(x_i(t), y_i(t))$ and $(w_i(t), h_i(t))$ be the
center and size of its box at time $t$, normalized by the frame width and
height. We define its dynamic state as:
\begin{equation}
\mathbf{z}_i(t) = \bigl[x_i(t),\, y_i(t),\, w_i(t),\, h_i(t),\,
\dot{x}_i(t),\, \dot{y}_i(t)\bigr]^{\top} \in \mathbb{R}^6,
\end{equation}
where the velocity $\mathbf{v}_i = [\dot{x}_i, \dot{y}_i]^{\top}$ is
estimated from the positions $\mathbf{p}_i = [x_i, y_i]^{\top}$ in
adjacent frames:
\begin{equation}
\mathbf{v}_i(t) = \frac{\mathbf{p}_i(t + \Delta t) - \mathbf{p}_i(t)}{\Delta t},
\end{equation}
with $\Delta t$ denoting the interval between adjacent frames.

\textbf{Position, Size, Velocity Similarity, and CLIP Score.}
For a reference trajectory $\tau^r_i$ and a generated trajectory
$\tau^g_j$, we compare their states at corresponding time steps $t_k$.
For brevity, we omit the indices $i$, $j$, and $t_k$ below.
The position similarity is based on the normalized center distance:
\begin{equation}
\label{eq:pos}
P_k = 1 - \frac{\|\mathbf{p}^r - \mathbf{p}^g\|_2}{\sqrt{2}}.
\end{equation}

The size similarity compares the width and height ratios:
\begin{equation}
\label{eq:size}
S_k = \sqrt{\frac{\min(w^r, w^g)}{\max(w^r, w^g)} \cdot
\frac{\min(h^r, h^g)}{\max(h^r, h^g)}}.
\end{equation}

Following the temporal correlation of
Animation2Code~\citep{ji2026animation2codeevaluatingtemporalvisual}, the velocity similarity
measures agreement in both motion direction and speed:

\begin{equation}
\label{eq:vel}
V_k = \max\!\left(0,\;
\frac{1}{2}\left(
\frac{\mathbf{v}^r \cdot \mathbf{v}^g}
{\|\mathbf{v}^r\|_2\,\|\mathbf{v}^g\|_2}
+
\frac{\min(\|\mathbf{v}^r\|_2,\|\mathbf{v}^g\|_2)}
{\max(\|\mathbf{v}^r\|_2,\|\mathbf{v}^g\|_2)}
\right)\right).
\end{equation}
We set $V_k = 1$ if both components are static and $0$ if only one is.

The CLIP score $C_k$~\citep{radford2021learning} compares the appearance of the two components on the image regions enclosed by their boxes. The pair scores $P_{ij}$, $S_{ij}$, $V_{ij}$, and $C_{ij}$ are obtained by averaging $P_k$, $S_k$, $V_k$, and $C_k$ over all time steps at which either component is present, where a time step with only one of them present scores zero.


\textbf{Trajectory Matching.}
We then match the reference and generated trajectories one to one (Stage~2
of Algorithm~\ref{alg:trajectory}). Trajectories whose OCR text
similarity $\mathrm{Text}(\tau^r_i, \tau^g_j)$ exceeds $\theta_t$ are
matched first. Following InTraGen~\citep{liu2024intragen}, the remaining
trajectories are matched by the Hungarian
algorithm~\citep{kuhn1955hungarian} under a cost $d_{ij}$ that combines
their position, size, velocity, and CLIP dissimilarities with weights $w_p=0.4$, $w_s=0.2$, $w_v=0.1$, and $w_c=0.3$.

\textbf{Score Aggregation.}
For segment $l$, let $\mathcal{T}^r_l$ and $\mathcal{T}^g_l$ be the
reference and generated trajectory sets, and let $\mathcal{A}_l$ be the
matched pairs. For each $Q \in \{P, S, V, C\}$, the segment score is
\begin{equation}
Q_l = \frac{\sum_{(i,j) \in \mathcal{A}_l} Q_{ij}}
{|\mathcal{T}^r_l| + |\mathcal{T}^g_l| - |\mathcal{A}_l|},
\label{eq:segment-score}
\end{equation}
So unmatched trajectories lower the score above. The final score averages
over all $L$ reference segments:
\begin{equation}
Q = \frac{1}{L} \sum_{l=1}^{L} Q_l.
\end{equation}

\section{Experiments}
\label{sec:experiments}

\begin{itemize}[leftmargin=*, itemsep=1pt, topsep=2pt]
    \item \textbf{(RQ1) Overall Performance.} How well do state-of-the-art MLLMs reconstruct animated webpages?
    \item \textbf{(RQ2) Coding Agents.} How effectively do coding agents reconstruct animated webpages, and what affects their performance?
    \item \textbf{(RQ3) Animation Types.} How does the performance of current systems vary across different animation types?
    \item \textbf{(RQ4) Human Evaluation.} How do humans assess the reconstructed animated webpages?
\end{itemize}

\begin{table*}[htbp]
\centering
\caption{Performance comparison. All metrics are higher-is-better. Models marked with $^\dagger$ receive the native video, and the others receive frames sampled at 2\,fps. The best and second-best scores are marked in bold and underlined, respectively.}
\label{tab:main-results}
\small
\setlength{\tabcolsep}{6pt}
\renewcommand{\arraystretch}{0.9}
\begin{tabular}{@{}lcccc@{\hspace{16pt}}cccc@{}}
\toprule
\multirow{2}{*}[-0.5ex]{\textbf{Model}} & \multicolumn{4}{c}{\textbf{Static Appearance Fidelity}} & \multicolumn{4}{c}{\textbf{Animation Fidelity}} \\
\cmidrule(lr){2-5}\cmidrule(l){6-9}
 & Structure $\uparrow$ & Text $\uparrow$ & Color $\uparrow$ & CLIP $\uparrow$ & Position $\uparrow$ & Size $\uparrow$ & Velocity $\uparrow$ & CLIP $\uparrow$ \\
\midrule
\rowcolor{gray!12}[0pt][0pt]
\multicolumn{9}{@{}c@{}}{\textit{Proprietary MLLMs}} \\
\hspace{0.8em}GPT-6 Astra~\cite{openai2026gpt6astra} & \textbf{0.6974} & \textbf{0.7804} & \textbf{0.7452} & \textbf{0.8796} & \textbf{0.4532} & \textbf{0.4193} & \textbf{0.3043} & \textbf{0.4303} \\
\hspace{0.8em}GPT-6 Sol~\cite{openai2026gpt6sol} & 0.5754 & 0.6726 & 0.6557 & 0.8084 & 0.3735 & 0.3359 & 0.2401 & 0.3511 \\
\hspace{0.8em}GPT-5.6 Sol~\cite{openai2026gpt56sol} & 0.5598 & 0.6503 & 0.6604 & 0.7898 & 0.3577 & 0.3164 & 0.2312 & 0.3355 \\
\hspace{0.8em}Claude Fable 5.1~\cite{anthropic2026fable51} & \underline{0.5952} & \underline{0.7101} & \underline{0.7014} & \underline{0.8516} & \underline{0.3928} & \underline{0.3495} & \underline{0.2653} & \underline{0.3660} \\
\hspace{0.8em}Claude Opus 5~\cite{anthropic2026opus5} & 0.5183 & 0.6240 & 0.6464 & 0.7907 & 0.3123 & 0.2710 & 0.2048 & 0.2922 \\
\hspace{0.8em}Grok 4.7~\cite{xai2026grok47} & 0.4575 & 0.6384 & 0.6169 & 0.7776 & 0.3170 & 0.2561 & 0.2133 & 0.2960 \\
\hspace{0.8em}Qwen3.8 Max (0902)$^\dagger$~\cite{qwen2026qwen38max} & 0.4273 & 0.5797 & 0.5589 & 0.7643 & 0.3016 & 0.2446 & 0.2026 & 0.2793 \\
\hspace{0.8em}Gemini 3.1 Pro Preview$^\dagger$~\cite{google2026gemini31pro} & 0.3453 & 0.5212 & 0.5449 & 0.7046 & 0.2313 & 0.1790 & 0.1456 & 0.2148 \\
\hspace{0.8em}Gemini 3.8 Flash$^\dagger$~\cite{google2026gemini38flash} & 0.3324 & 0.5061 & 0.5220 & 0.6794 & 0.2301 & 0.1761 & 0.1530 & 0.2117 \\
\hspace{0.8em}Doubao-Seed-2.1-pro$^\dagger$~\cite{bytedance2026seed21} & 0.3378 & 0.4695 & 0.5149 & 0.6127 & 0.2060 & 0.1586 & 0.1238 & 0.1908 \\
\midrule
\rowcolor{gray!12}[0pt][0pt]
\multicolumn{9}{@{}c@{}}{\textit{Open-Weight MLLMs}} \\
\hspace{0.8em}Kimi K3$^\dagger$~\cite{moonshot2026kimik3} & 0.5373 & 0.6775 & 0.6564 & 0.8204 & 0.3463 & 0.2978 & 0.2246 & 0.3218 \\
\hspace{0.8em}DeepSeek V4.1 Flash~\cite{deepseek2026v41flash} & 0.4393 & 0.5820 & 0.5854 & 0.7530 & 0.3063 & 0.2543 & 0.1983 & 0.2836 \\
\hspace{0.8em}GLM 5.3 Flash$^\dagger$~\cite{zhipu2026glm53flash} & 0.4293 & 0.5707 & 0.5750 & 0.7441 & 0.2953 & 0.2390 & 0.1960 & 0.2733 \\
\hspace{0.8em}Qwen3 VL 235B A22B Thinking$^\dagger$~\cite{bai2025qwen3vl} & 0.2395 & 0.4032 & 0.4337 & 0.5770 & 0.1702 & 0.1294 & 0.1170 & 0.1584 \\
\bottomrule
\end{tabular}
\end{table*}

\subsection{Experimental Setup}

\textbf{Models.}
We evaluate 14 MLLMs on AnimWeb-Bench, including 10 proprietary and 4 open-weight models, as listed in Table~\ref{tab:main-results}. For each task, models are provided with the reference video and a set of composite images containing all processed static assets of the original webpage.
Models with native video support (marked with $^\dagger$ in Table~\ref{tab:main-results}) take the reference video directly.
To accommodate models without native video support, we uniformly sample frames from the video at 2 frames per second and input them as an image sequence.
The task is to generate a single self-contained HTML file with inline CSS and JavaScript that reproduces both the appearance and the animations of the reference webpage.
Each generated page is then rendered and recorded using the same configuration as the reference videos, capturing its own animation sequence in full or one complete cycle if it loops.
All other inference parameters follow the default settings recommended by the model providers.

\textbf{Coding Agents.}
Beyond single-pass generation, we further evaluate 3 coding agents, each powered by one of the MLLMs above: Codex~\cite{openai2026codex} with GPT-5.6 Sol, Claude Code~\cite{anthropic2026claudecode} with Claude Opus 5, and Qwen Code~\cite{qwen2026qwencode} with Qwen3.8 Max (0902), all with medium reasoning effort.
Each agent operates in a sandboxed workspace that contains the task inputs, where it can execute commands, create and edit files, and inspect the inputs, but cannot access outside this workspace or the network.
Unlike MLLMs, which receive inputs directly in their context, agents access the inputs as files and can decide how to process them, for example by extracting frames from the video.
To examine how input format affects performance, we evaluate each agent under two settings: \emph{Video}, which provides the reference video, and \emph{Frames}, which provides the same 2 fps frames given to the MLLMs.
In both settings, the workspace contains the composite asset images. The agent returns a single HTML file, which is evaluated in the same way as the MLLM outputs.

\textbf{Human Evaluation.}
While automatic metrics provide a fine-grained assessment of model performance, evaluating how users perceive the generated webpages is also essential.
Following Design2Code~\cite{si2025design2code}, we conduct a pairwise human evaluation.
(1) \emph{Participants.} We recruit five annotators through a local university mailing list. All annotators hold at least a B.S. degree in computer science or software engineering and have more than two years of web-development experience.
(2) \emph{Procedure.} We randomly sample 50 webpages from AnimWeb-Bench and use Qwen3.8 Max (0902) as the baseline.
We select eight models and pair the output of each with that of the baseline on every sampled webpage. Throughout the evaluation, annotators follow a guideline~\cite{animwebbench_human_evaluation_guideline} that defines the evaluation criteria for static appearance and animation and provides instructions for handling ambiguous cases.
Annotators view the reference video together with the recordings of the two generated pages, and judge each pair separately on static appearance and animation by selecting ``Example 1 better'', ``Example 2 better'', or ``Tie''.
Annotators are blind to which model produced each page, and pairs are presented in randomized order.
The final label of each pair is determined by majority voting among the five annotators.

\subsection{RQ1: Overall Performance}
\label{sec:rq1}
Table~\ref{tab:main-results} reports the results of the 14 MLLMs on AnimWeb-Bench.

\begin{figure*}[htbp]
  \centering
  \includegraphics[width=\textwidth]{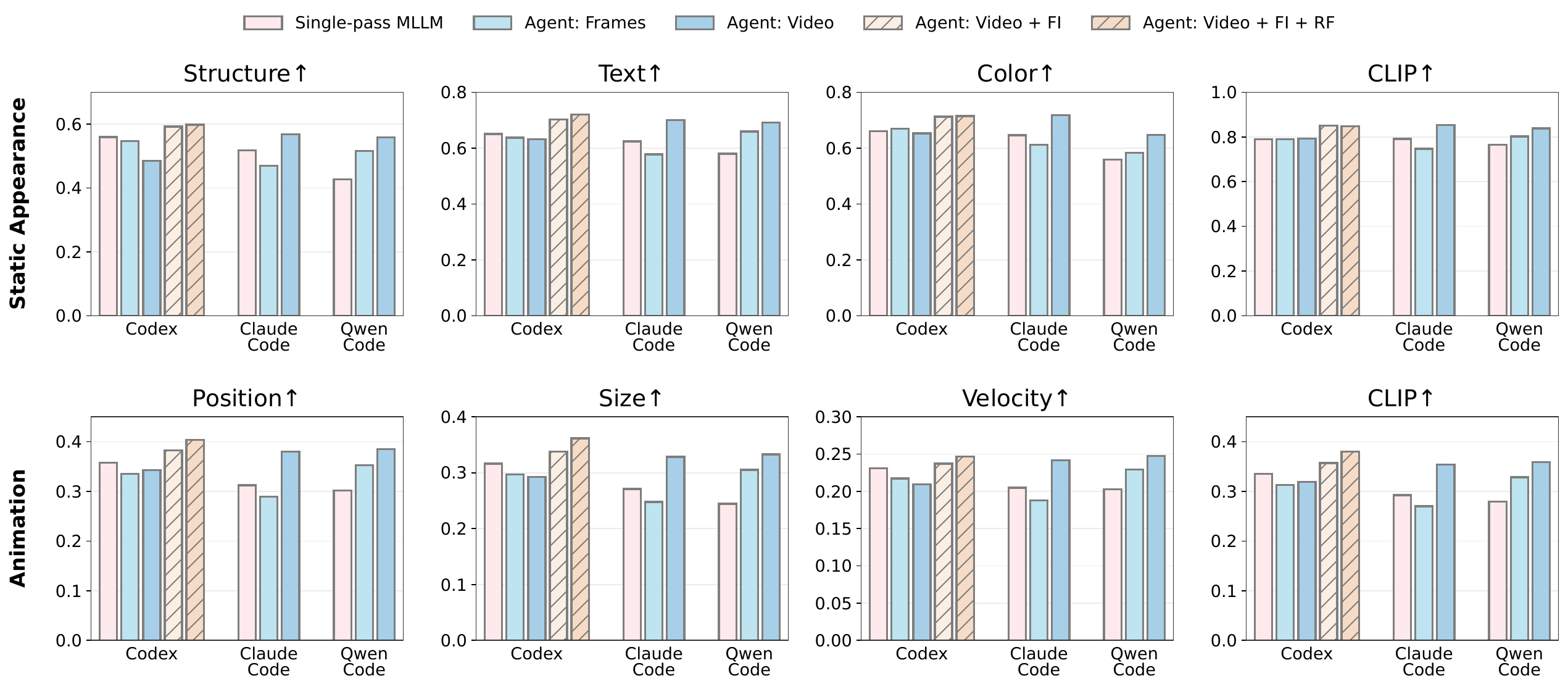}
  \caption{Comparison of coding agents with their underlying MLLMs in single-pass generation. Codex is additionally evaluated with frame inspection (FI) and render feedback (RF).}
  \label{fig:rq2-agents}
\end{figure*}



\textbf{Systems Look Right but Move Wrong.}
Animation fidelity is low across the board.
Even GPT-6 Astra, the strongest model, scores below 0.46 on every animation metric, and ten of the remaining thirteen models stay below 0.35.
For every model, each animation score falls below each static score, and Velocity is the lowest of all eight metrics.
Because the two dimensions are computed differently, their absolute values are not directly comparable.
Two further observations, however, show that the gap reflects motion rather than scale.
Animation metrics respond to the kind of motion on a page while static metrics do not (Section~\ref{sec:rq3}), and human annotators prefer every evaluated model over the baseline less often on animation than on static appearance (Section~\ref{sec:rq4}).

Current models fail to infer and encode component-level dynamics in three ways.
First, some animated components are missing from the generated page altogether, and their unmatched trajectories lower all four animation scores (Eq.~\eqref{eq:segment-score}).
Second, even when a component is animated, its trajectory or manner of change may deviate from the reference. For example, sliding along a different path or scaling instead of sliding, which lowers Position and Size.
Third, and most subtle, is the temporal behavior of motion, namely when and how fast each component moves.
Components may move faster, slower, or in a different direction than in the reference, which lowers Velocity.
They may also appear at the wrong time: elements that the reference reveals gradually or one after another may appear all at once in the generated page, leaving time steps at which only one of the matched components is present and thus lowering all four animation scores.
Appendix~\ref{app:case_study} illustrates each of these failures on two webpages.
Faithfully recovering how a webpage moves thus remains an open challenge for current models.

\textbf{Proprietary Models Lead, but Open-Weight Models Are Closing In.}
GPT-6 Astra leads all eight metrics, and Claude Fable 5.1 ranks second on every metric.
Among open-weight models, Kimi K3 ranks third in static Text and CLIP, surpassing GPT-6 Sol, and fifth across all animation metrics, outperforming Claude Opus 5 and approaching GPT-5.6 Sol.
DeepSeek V4.1 Flash and GLM 5.3 Flash perform on par with the proprietary Grok 4.7 and Qwen3.8 Max (0902), and all four outperform the two Gemini models and Doubao-Seed-2.1-pro across all metrics.

\subsection{RQ2: Coding Agents}
\label{sec:rq2}

Figure~\ref{fig:rq2-agents} compares coding agents with their single-pass MLLMs.


\textbf{Coding Agents Do Not Consistently Outperform Their Underlying Models.}
We compare each agent with its own underlying MLLM, and find that the benefit of agentic generation depends on both the agent and the input format.
Qwen Code outperforms Qwen3.8 Max (0902) on all eight metrics under both settings, with relative gains of 4.3\%--24.7\% under Frames and 9.7\%--36.1\% under Video.
Claude Code shows a reversal: it falls below Claude Opus 5 on all eight metrics under Frames, yet surpasses it on all eight under Video, with relative gains of 7.9\%--21.7\%.
Codex, by contrast, trails GPT-5.6 Sol on six metrics under Frames and seven under Video, with the remaining metrics changing by at most 1.4\%.
Comparing the two input formats, Video yields higher scores than Frames on all eight metrics for both Claude Code and Qwen Code, suggesting that video is generally the more informative input.
Unlike the fixed 2 fps frames, a video allows an agent to choose its own sampling rate, for example, sampling densely around transitions, and can thus reveal intermediate states and timing that sparse frames miss.
For Codex, however, Video outperforms Frames on only three metrics, indicating that this flexibility does not automatically translate into better reconstruction.
The way in which an agent perceives visual input is therefore critical to its performance.

\textbf{Frame Inspection and Render Feedback Improve Agentic Reconstruction.}
To understand why Codex lags behind, we examine its execution traces under Video.
We observe that Codex often extracts frames from the video, tiles them into contact sheets (e.g., nine frames per image), and relies on these sheets rather than on the individual frames (Appendix~\ref{app:trace}).
Such reduced tiles may obscure fine visual details and subtle changes between frames.
We therefore introduce two additional Video settings for Codex.
In \emph{Video + Frame Inspection} (Video+FI), the prompt instructs Codex to also inspect each source frame individually at its original resolution whenever it builds a contact sheet.
\emph{Video + Frame Inspection + Render Feedback} (Video+FI+RF) further provides a recording tool that renders the generated page into a video, allowing Codex to compare its output with the reference and revise the code iteratively.
The two enhancements improve Codex in different ways.
Compared with Video, Video+FI improves all eight metrics by 7.3\%--22.0\% on the static metrics and 11.6\%--15.6\% on the animation metrics, and outperforms GPT-5.6 Sol on every metric.
Building on Video+FI, Video+FI+RF further raises all four animation metrics by 4.0\%--7.0\%, whereas the static metrics change by only $-0.3\%$ to $2.5\%$.
Frame inspection thus improves overall fidelity, while render feedback mainly benefits animation, likely because comparing the rendered page with the reference video exposes motion errors that are hard to detect from the code alone.
With both enhancements, Codex outperforms GPT-6 Sol on all eight metrics and Claude Fable 5.1 on six, second only to GPT-6 Astra on most metrics.
These results suggest that the bottleneck of agentic reconstruction lies not only in code generation but also in perception and verification: an agent improves substantially once it observes the reference closely and checks how its own page actually moves.
Because the frame-inspection instruction was derived from Codex's traces on this benchmark, we treat Video+FI and Video+FI+RF as diagnostic interventions that locate the bottleneck rather than as tuned methods.

\begin{table}[t!]
\centering
\caption{Average scores of the 14 MLLMs on webpages grouped by animation category and temporal behavior.}
\label{tab:animation-groups}
\setlength{\tabcolsep}{3pt}
\resizebox{\columnwidth}{!}{%
\begin{tabular}{@{}lcccc@{\hspace{8pt}}cccc@{}}
\toprule
\multirow{2}{*}[-0.5ex]{\textbf{Group}}
 & \multicolumn{4}{c}{\textbf{Static Appearance}}
 & \multicolumn{4}{c}{\textbf{Animation}} \\
\cmidrule(lr){2-5}\cmidrule(l){6-9}
 & Struct. & Text & Color & CLIP & Pos. & Size & Vel. & CLIP \\
\midrule
\multicolumn{9}{@{}l}{\textit{By Animation Category}} \\
Spatial Motion  & 0.4615 & 0.6110 & 0.6049 & 0.7523 & 0.2911 & 0.2445 & 0.1855 & 0.2703 \\
Property Change & 0.4654 & 0.5702 & 0.5919 & 0.7505 & 0.3190 & 0.2701 & 0.2141 & 0.2988 \\
Visibility Only & 0.4811 & 0.6715 & 0.6284 & 0.8056 & 0.4325 & 0.3797 & 0.3291 & 0.4095 \\
\midrule
\multicolumn{9}{@{}l}{\textit{By Temporal Behavior}} \\
Rapid, Continuous Change & 0.3761 & 0.4972 & 0.5299 & 0.6883 & 0.2379 & 0.1969 & 0.1307 & 0.2228 \\
Discrete State Transition      & 0.4886 & 0.6305 & 0.6414 & 0.7761 & 0.4194 & 0.3591 & 0.3004 & 0.3940 \\
\bottomrule
\end{tabular}%
}
\end{table}

\subsection{RQ3: Animation Types}
\label{sec:rq3}

To compare performance across animation types, we divide the 124 webpages into three mutually exclusive groups based on animation type labels.
(1) \emph{Spatial Motion} contains at least one animation that moves components in space, such as translation, rotation, shake, or card flip.
(2) \emph{Property Change} contains no spatial motion but includes changes beyond visibility, such as scaling, color transitions, text animations, or numeric counters.
(3) \emph{Visibility Only} contains only animations in which components appear or fade in.
The top part of Table~\ref{tab:animation-groups} reports the average scores of the 14 MLLMs on each group.

\textbf{Spatial Motion Is the Hardest to Reproduce.}
Animation fidelity declines as the animations involve more motion.
On animation metrics, Visibility Only scores the highest, followed by Property Change, while Spatial Motion scores the lowest, with a Velocity only 56\% that of Visibility Only.
This ordering reflects the demands of each group.
Visibility changes only require models to determine when elements appear, and property changes alter the size, color, or content of components with limited motion, whereas spatial motion requires recovering how each element travels over time.
The static metrics behave differently: between Spatial Motion and Property Change, they differ by $-0.8\%$ to $+7.2\%$ with no consistent direction, whereas all four animation metrics drop by 8.7\%--13.4\%.
Animation categories thus mainly affect how well models reproduce motion rather than appearance, which shows that Animation Fidelity captures differences that the static metrics miss.

\textbf{Rapid, Continuous Changes Are Harder than Discrete State Transitions.}
Among individual animation types, the most challenging ones change the page rapidly and continuously, often across many elements at once, such as Card Flip, Twinkling, Text Scramble, Text Fragmentation, Shake, and Numeric Counter Animation.
Their intermediate states shift at nearly every frame, making the timing difficult to infer from the reference video and to express precisely in code.
The easiest types, such as Font Weight Animation, Icon State Transition, and Text Substitution, are instead discrete transitions between two clearly visible states, requiring models only to identify the start and end states and connect them with a simple transition. This contrast is also reflected in the bottom part of Table~\ref{tab:animation-groups}, where webpages with rapid, continuous changes score lower on all metrics than those with discrete state transitions.

\subsection{RQ4: Human Evaluation}
\label{sec:rq4}

\begin{figure}[t]
    \centering
    \includegraphics[width=0.48\textwidth]{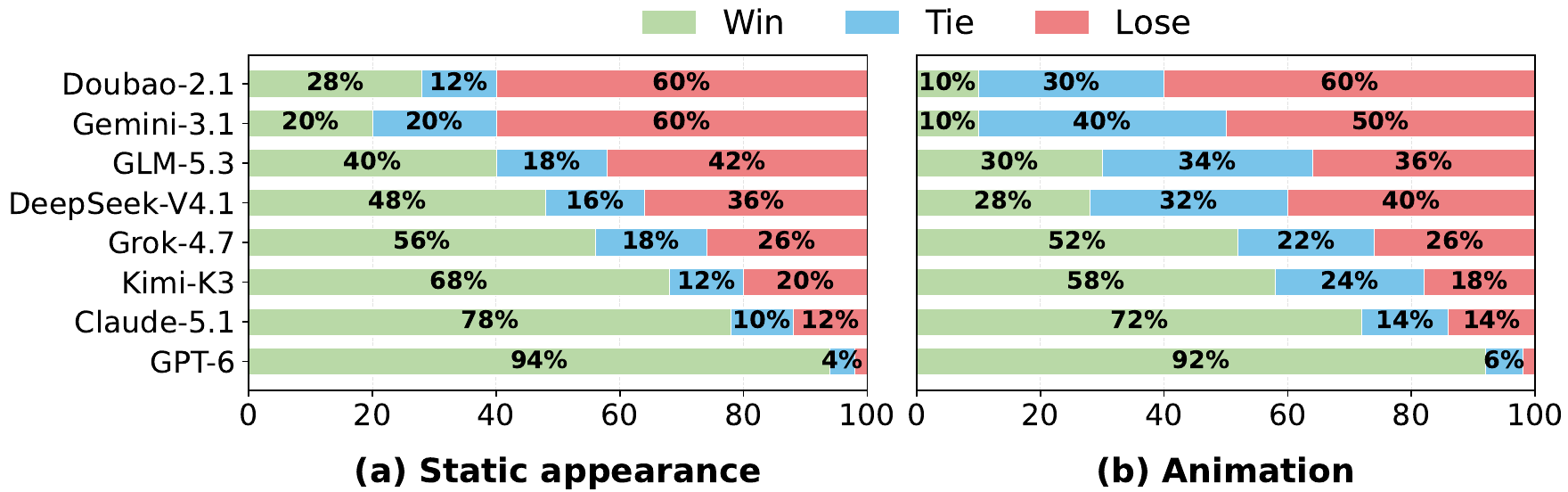} 
    \caption{Human evaluation results against Qwen3.8 Max (0902) on 50 sampled webpages.}
   \label{fig:human-eval}
\end{figure}

Figure~\ref{fig:human-eval} shows the human evaluation results, where higher win rates and lower loss rates indicate better reconstruction. 

\textbf{Human Judgments Align with Automatic Metrics.}
In both dimensions, human evaluators rank the models in nearly the same order as our automatic metrics.
GPT-6 Astra is preferred over the baseline on 94\% of the webpages for static appearance and 92\% for animation, followed by Claude Fable 5.1, Kimi K3, and Grok 4.7, whose win rates decline in the same order as their automatic scores.
DeepSeek V4.1 Flash and GLM 5.3 Flash perform close to the baseline, with comparable win and loss rates, consistent with their automatic scores differing from those of Qwen3.8 Max (0902) by less than 0.03 on every metric.
Gemini 3.1 Pro Preview and Doubao-Seed-2.1-pro lose to the baseline on at least half of the webpages in both dimensions, again matching their low automatic scores. All eight metrics correlate positively with human preferences, with Spearman's $\rho$ ranging from 0.667 to 0.737 (Appendix~\ref{app:human-correlation}).

\textbf{Animations Are Harder for Humans to Distinguish.}
Compared with static appearance, every model obtains a lower win rate and a higher tie rate on animation.
The rise in ties is particularly pronounced for weaker models: the tie rates of DeepSeek V4.1 Flash, GLM 5.3 Flash, Gemini 3.1 Pro Preview, and Doubao-Seed-2.1-pro rise from 12\%--20\% to 30\%--40\%.

When neither animation closely follows the reference, annotators find little basis to prefer one over the other, echoing our finding in Section~\ref{sec:rq1} that reproducing animation remains challenging.

\section{Conclusion}

We presented AnimWeb-Bench, a benchmark for animated webpage code generation from videos, together with a human-aligned evaluation protocol that measures both static appearance and animation fidelity. Our systematic evaluation of MLLMs and coding agents shows that while current systems reproduce static appearance reasonably well, they still struggle to capture the temporal dynamics of webpage animations. Our experiments further demonstrate the benefits of careful visual inspection and render feedback for coding agents. We hope AnimWeb-Bench will support future research on temporal visual understanding and code generation toward more faithful reconstruction of animated webpages.

\bibliographystyle{ACM-Reference-Format}
\bibliography{references}

@misc{animwebbench_human_evaluation_guideline,
  title  = {AnimWeb-Bench Human Evaluation Guidelines},
  year   = {2026},
  url    = {https://github.com/starry1001/AnimWeb/blob/main/docs/human_evaluation_guideline.md},
  note   = {Accessed: 2026-11-01}
}

@misc{laurencon2024unlocking,
      title={Unlocking the conversion of Web Screenshots into HTML Code with the WebSight Dataset}, 
      author={Hugo Laurençon and Léo Tronchon and Victor Sanh},
      year={2024},
      eprint={2403.09029},
      archivePrefix={arXiv},
      doi={10.48550/arXiv.2403.09029},
      primaryClass={cs.HC}
}

@inproceedings{yun2024web2code,
author = {Yun, Sukmin and Lin, Haokun and Thushara, Rusiru and Bhat, Mohammad Qazim and Wang, Yongxin and Jiang, Zutao and Deng, Mingkai and Wang, Jinhong and Tao, Tianhua and Li, Junbo and Li, Haonan and Nakov, Preslav and Baldwin, Timothy and Liu, Zhengzhong and Xing, Eric P. and Liang, Xiaodan and Shen, Zhiqiang},
title = {Web2Code: a large-scale webpage-to-code dataset and evaluation framework for multimodal LLMs},
year = {2024},
isbn = {9798331314385},
publisher = {Curran Associates Inc.},
booktitle = {Proceedings of the 38th International Conference on Neural Information Processing Systems},
articleno = {3560},
numpages = {24},
location = {Vancouver, BC, Canada},
series = {NeurIPS '24}
}

@inproceedings{si2025design2code,
    title = "{D}esign2{C}ode: Benchmarking Multimodal Code Generation for Automated Front-End Engineering",
    author = "Si, Chenglei  and
      Zhang, Yanzhe  and
      Li, Ryan  and
      Yang, Zhengyuan  and
      Liu, Ruibo  and
      Yang, Diyi",
    editor = "Chiruzzo, Luis  and
      Ritter, Alan  and
      Wang, Lu",
    booktitle = "Proceedings of the 2025 Conference of the Nations of the Americas Chapter of the Association for Computational Linguistics: Human Language Technologies (Volume 1: Long Papers)",
    month = apr,
    year = "2025",
    address = "Albuquerque, New Mexico",
    publisher = "Association for Computational Linguistics",
    url = "https://aclanthology.org/2025.naacl-long.199/",
    doi = "10.18653/v1/2025.naacl-long.199",
    pages = "3956--3974",
    ISBN = "979-8-89176-189-6"
}

@inproceedings{gui2025webcode2m,
author = {Gui, Yi and Li, Zhen and Wan, Yao and Shi, Yemin and Zhang, Hongyu and Chen, Bohua and Su, Yi and Chen, Dongping and Wu, Siyuan and Zhou, Xing and Jiang, Wenbin and Jin, Hai and Zhang, Xiangliang},
title = {WebCode2M: A Real-World Dataset for Code Generation from Webpage Designs},
year = {2025},
isbn = {9798400712746},
publisher = {Association for Computing Machinery},
address = {New York, NY, USA},
url = {https://doi.org/10.1145/3696410.3714889},
doi = {10.1145/3696410.3714889},
booktitle = {Proceedings of the ACM on Web Conference 2025},
pages = {1834–1845},
numpages = {12},
location = {Sydney NSW, Australia},
series = {WWW '25}
}

@inproceedings{xiao2024interaction2code,
author = {Xiao, Jingyu and Wan, Yuxuan and Huo, Yintong and Wang, Zixin and Xu, Xinyi and Wang, Wenxuan and Xu, Zhiyao and Wang, Yuhang and Lyu, Michael R.},
title = {Interaction2Code: Benchmarking MLLM-based Interactive Webpage Code Generation from Interactive Prototyping},
year = {2025},
publisher = {IEEE Press},
url = {https://doi.org/10.1109/ASE63991.2025.00028},
doi = {10.1109/ASE63991.2025.00028},
booktitle = {2025 40th IEEE/ACM International Conference on Automated Software Engineering (ASE)},
pages = {241–253},
numpages = {13},
location = {Seoul, Korea, Republic of}
}

@inproceedings{chen2026iwr,
  title={IWR-Bench: Can LVLMs reconstruct interactive webpage from a user interaction video?},
  author={Chen, Yang and Liu, Minghao and Shen, Yufan and Li, Yunwen and Huang, Tianyuan and Fang, Xinyu and Zheng, Tianyu and Huang, Wenxuan and Yang, Cheng and Wen, Licheng and others},
  booktitle={International Conference on Learning Representations},
  volume={2026},
  pages={52158--52189},
  year={2026},
  url={https://arxiv.org/abs/2509.24709}
}

@misc{dai2026webvrbenchmarkingmultimodalllms,
      title={WebVR: Benchmarking Multimodal LLMs for WebPage Recreation from Videos via Human-Aligned Visual Rubrics}, 
      author={Yuhong Dai and Yanlin Lai and Mitt Huang and Hangyu Guo and Dingming Li and Hongbo Peng and Haodong Li and Yingxiu Zhao and Haoran Lyu and Zheng Ge and Xiangyu Zhang and Daxin Jiang},
      year={2026},
      eprint={2603.13391},
      archivePrefix={arXiv},
      primaryClass={cs.CV},
      url={https://arxiv.org/abs/2603.13391}, 
}

@misc{ji2026animation2codeevaluatingtemporalvisual,
      title={Animation2Code: Evaluating Temporal Visual Reasoning in Video-to-Code Generation}, 
      author={Anya Ji and Abhijith Varma Mudunuri and David M. Chan and Alane Suhr},
      year={2026},
      eprint={2606.28593},
      archivePrefix={arXiv},
      primaryClass={cs.CV},
      url={https://arxiv.org/abs/2606.28593}, 
}

@inproceedings{wan2026TDDev,
author = {Wan, Yuxuan and Liang, Tingshuo and Xu, Jiakai and Xiao, Jingyu and Huo, Yintong and Lyu, Michael R.},
title = {From Runnable Code to Shippable Applications: Test-Driven Development for Full-Stack Web Application Generation},
year = {2026},
isbn = {9798400728822},
publisher = {Association for Computing Machinery},
address = {New York, NY, USA},
url = {https://doi.org/10.1145/3832783.3834388},
doi = {10.1145/3832783.3834388},
booktitle = {Proceedings of the 41st IEEE/ACM International Conference on Automated Software Engineering},
pages = {726–738},
numpages = {13},
location = {Munich, Germany},
series = {ASE '26}
}

@article{xiao2026visualrepair,
  title={VisualRepair: Dynamic Tool Calling and Region Focusing for Visual Software Issue Repair},
  author={Xiao, Jingyu and Zhang, Zhongyi and Hou, Haoran and Wan, Yuxuan and Jiang, Yuan and Huo, Yintong and Lyu, Michael R},
  journal={arXiv preprint arXiv:2607.14075},
  year={2026}
}

@article{zhao2026beyond,
  title={Beyond NL2Code: A Structured Survey of Multimodal Code Intelligence},
  author={Zhao, Xuanle and Sun, Qiushi and Xiao, Jingyu and Liu, Xuexin and Yang, Haoyue and Chen, Qiaosheng and Luo, Xianzhen and Huang, Jing and Zhong, Yufeng and Chen, Lei and others},
  journal={arXiv preprint arXiv:2606.15932},
  year={2026}
}

@inproceedings{dang2025envisioning,
  title={Envisioning Future Interactive Web Development: Editing Webpage with Natural Language},
  author={Dang, Truong Hai and Xiao, Jingyu and Huo, Yintong},
  booktitle={2025 2nd IEEE/ACM International Conference on AI-powered Software (AIware)},
  pages={61--66},
  year={2025},
  organization={IEEE}
}

@inproceedings{le2026uibenchkit,
  title={UIBenchKit: A unified toolkit for design-to-code model evaluation},
  author={Le, Chinh T and Ong, Trevor Yee Siang and Xiao, Jingyu and Wan, Yuxuan and Huo, Yintong},
  booktitle={Proceedings of the 41st IEEE/ACM International Conference on Automated Software Engineering},
  pages={3445--3449},
  year={2026}
}

@article{wan2024mrweb,
  title={MRWeb: An Exploration of Generating Multi-Page Resource-Aware Web Code from UI Designs},
  author={Wan, Yuxuan and Dong, Yi and Xiao, Jingyu and Huo, Yintong and Wang, Wenxuan and Lyu, Michael R},
  journal={arXiv preprint arXiv:2412.15310},
  doi={10.48550/arXiv.2412.15310},
  year={2024}
}

@article{xiao2025designbench,
  title={Designbench: A comprehensive benchmark for mllm-based front-end code generation},
  author={Xiao, Jingyu and Wang, Ming and Lam, Man Ho and Wan, Yuxuan and Liu, Junliang and Huo, Yintong and Lyu, Michael R},
  journal={arXiv preprint arXiv:2506.06251},
  doi = {10.48550/arXiv.2506.06251},
  year={2025}
}

@inproceedings{radford2021learning,
  title={Learning transferable visual models from natural language supervision},
  author={Radford, Alec and Kim, Jong Wook and Hallacy, Chris and Ramesh, Aditya and Goh, Gabriel and Agarwal, Sandhini and Sastry, Girish and Askell, Amanda and Mishkin, Pamela and Clark, Jack and others},
  booktitle={International conference on machine learning},
  pages={8748--8763},
  year={2021},
  organization={PmLR}
}

@article{levenshtein1966binary,
  title   = {Binary Codes Capable of Correcting Deletions, Insertions, and Reversals},
  author  = {Levenshtein, Vladimir I.},
  journal = {Soviet Physics Doklady},
  volume  = {10},
  number  = {8},
  pages   = {707--710},
  year    = {1966}
}

@inproceedings{xiao2026component,
  title={Component-based Reusable UI Code Generation for Complex Websites via Semantic Segmentation and Fine-grained Feedback},
  author={Xiao, Jingyu and Qin, Jiantong and Li, Shuoqi and Lam, Man Ho and Wan, Yuxuan and Huang, Jen-tse and Huo, Yintong and Lyu, Michael R},
  booktitle={Proceedings of the 32nd ACM SIGKDD Conference on Knowledge Discovery and Data Mining V. 2},
  pages={5638--5649},
  year={2026}
}

@article{luo2001development,
  title={The development of the CIE 2000 colour-difference formula: CIEDE2000},
  author={Luo, M Ronnier and Cui, Guihua and Rigg, Bryan},
  journal={Color Research \& Application: Endorsed by Inter-Society Color Council, The Colour Group (Great Britain), Canadian Society for Color, Color Science Association of Japan, Dutch Society for the Study of Color, The Swedish Colour Centre Foundation, Colour Society of Australia, Centre Fran{\c{c}}ais de la Couleur},
  volume={26},
  number={5},
  pages={340--350},
  year={2001},
  publisher={Wiley Online Library}
}

@misc{salesforce2026gpa,
  author       = {{Salesforce AI Research}},
  title        = {{GPA-GUI-Detector}},
  year         = {2026},
  howpublished = {\url{https://huggingface.co/Salesforce/GPA-GUI-Detector}}
}

@article{liu2024intragen,
  title={InTraGen: Trajectory-controlled video generation for object interactions},
  author={Liu, Zuhao and Yanev, Aleksandar and Mahmood, Ahmad and Nikolov, Ivan and Motamed, Saman and Zheng, Wei-Shi and Wang, Xi and Sun, Lei and Van Gool, Luc and Paudel, Danda Pani},
  journal={arXiv preprint arXiv:2411.16804},
  year={2024}
}

@article{kuhn1955hungarian,
  title={The Hungarian method for the assignment problem},
  author={Kuhn, Harold W},
  journal={Naval research logistics quarterly},
  volume={2},
  number={1-2},
  pages={83--97},
  year={1955},
  publisher={Wiley Online Library}
}

@inproceedings{gui2025latcoder,
  title={Latcoder: Converting webpage design to code with layout-as-thought},
  author={Gui, Yi and Li, Zhen and Zhang, Zhongyi and Wang, Guohao and Lv, Tianpeng and Jiang, Gaoyang and Liu, Yi and Chen, Dongping and Wan, Yao and Zhang, Hongyu and others},
  booktitle={Proceedings of the 31st ACM SIGKDD Conference on Knowledge Discovery and Data Mining V. 2},
  pages={721--732},
  year={2025}
}

@inproceedings{beltramelli2018pix2code,
  title={pix2code: Generating code from a graphical user interface screenshot},
  author={Beltramelli, Tony},
  booktitle={Proceedings of the ACM SIGCHI symposium on engineering interactive computing systems},
  pages={1--6},
  year={2018}
}

@article{jiang2025screencoder,
  title={Screencoder: Advancing visual-to-code generation for front-end automation via modular multimodal agents},
  author={Jiang, Yilei and Zheng, Yaozhi and Wan, Yuxuan and Han, Jiaming and Wang, Qunzhong and Lyu, Michael R and Yue, Xiangyu},
  journal={arXiv preprint arXiv:2507.22827},
  year={2025}
}

@misc{openai2026gpt6astra,
  author = {{OpenAI}},
  title  = {{GPT-6 Astra}: A New Generation of Intelligence},
  year   = {2026},
  month  = sep,
  url    = {https://openai.com/index/gpt-6-astra/}
}

@misc{openai2026gpt6sol,
  author = {{OpenAI}},
  title  = {Introducing {GPT-6 Sol} and {Luna}},
  year   = {2026},
  month  = sep,
  url    = {https://openai.com/index/introducing-gpt-6-sol-and-luna/}
}

@misc{openai2026gpt56sol,
  author = {{OpenAI}},
  title  = {Previewing {GPT-5.6 Sol}: A Next-Generation Model},
  year   = {2026},
  month  = jun,
  url    = {https://openai.com/index/previewing-gpt-5-6-sol/}
}

@misc{anthropic2026fable51,
  author = {{Anthropic}},
  title  = {{Claude Fable 5.1} and {Claude Mythos 5.1}},
  year   = {2026},
  month  = sep,
  url    = {https://www.anthropic.com/claude-fable-and-mythos-5-1}
}

@misc{anthropic2026opus5,
  author = {{Anthropic}},
  title  = {Introducing {Claude Opus 5}},
  year   = {2026},
  month  = jul,
  url    = {https://www.anthropic.com/news/claude-opus-5}
}

@misc{xai2026grok47,
  author = {{SpaceXAI}},
  title  = {Introducing {Grok 4.7}},
  year   = {2026},
  month  = sep,
  url    = {https://x.ai/news/grok-4-7}
}

@misc{qwen2026qwen38max,
  author = {{Qwen Team}},
  title  = {{Qwen3.8-Max}: A New Bar for Coding and Cowork},
  year   = {2026},
  month  = aug,
  url    = {https://qwen.ai/blog?id=qwen3.8}
}

@misc{google2026gemini31pro,
  author = {{Google}},
  title  = {{Gemini 3.1 Pro}},
  year   = {2026},
  month  = feb,
  url    = {https://blog.google/innovation-and-ai/models-and-research/gemini-models/gemini-3-1-pro/}
}

@misc{google2026gemini38flash,
  author = {{Google}},
  title  = {{Gemini 3.8 Flash} and {Gemini 3.8 Flash Cyber}},
  year   = {2026},
  month  = sep,
  url    = {https://blog.google/innovation-and-ai/models-and-research/gemini-models/3-8-flash-and-3-8-flash-cyber/}
}

@misc{bytedance2026seed21,
  author = {{ByteDance Seed}},
  title  = {{Seed2.1} Officially Released: Advancing {AI} Productivity},
  year   = {2026},
  month  = jun,
  url    = {https://seed.bytedance.com/en/blog/seed2-1-officially-released-advancing-ai-productivity}
}

@techreport{moonshot2026kimik3,
  author      = {{Kimi Team}},
  title       = {{Kimi K3}: Open Frontier Intelligence},
  institution = {Moonshot AI},
  year        = {2026},
  month       = jul,
  url         = {https://github.com/MoonshotAI/Kimi-K3}
}

@misc{deepseek2026v41flash,
  author = {{DeepSeek-AI}},
  title  = {{DeepSeek-V4.1-Flash}},
  year   = {2026},
  month  = sep,
  url    = {https://huggingface.co/deepseek-ai/DeepSeek-V4.1-Flash}
}

@misc{zhipu2026glm53flash,
  author = {{Z.ai}},
  title  = {{GLM-5.3-Flash}},
  year   = {2026},
  month  = aug,
  url    = {https://huggingface.co/zai-org/GLM-5.3-Flash}
}

@misc{bai2025qwen3vl,
      title={Qwen3-VL Technical Report}, 
      author={Shuai Bai and Yuxuan Cai and Ruizhe Chen and Keqin Chen and Xionghui Chen and Zesen Cheng and Lianghao Deng and Wei Ding and Chang Gao and Chunjiang Ge and Wenbin Ge and Zhifang Guo and Qidong Huang and Jie Huang and Fei Huang and Binyuan Hui and Shutong Jiang and Zhaohai Li and Mingsheng Li and Mei Li and Kaixin Li and Zicheng Lin and Junyang Lin and Xuejing Liu and Jiawei Liu and Chenglong Liu and Yang Liu and Dayiheng Liu and Shixuan Liu and Dunjie Lu and Ruilin Luo and Chenxu Lv and Rui Men and Lingchen Meng and Xuancheng Ren and Xingzhang Ren and Sibo Song and Yuchong Sun and Jun Tang and Jianhong Tu and Jianqiang Wan and Peng Wang and Pengfei Wang and Qiuyue Wang and Yuxuan Wang and Tianbao Xie and Yiheng Xu and Haiyang Xu and Jin Xu and Zhibo Yang and Mingkun Yang and Jianxin Yang and An Yang and Bowen Yu and Fei Zhang and Hang Zhang and Xi Zhang and Bo Zheng and Humen Zhong and Jingren Zhou and Fan Zhou and Jing Zhou and Yuanzhi Zhu and Ke Zhu},
      year={2025},
      eprint={2511.21631},
      archivePrefix={arXiv},
      primaryClass={cs.CV},
      url={https://arxiv.org/abs/2511.21631}, 
}

@misc{openai2026codex,
  author = {{OpenAI}},
  title  = {{Codex}: A Lightweight Coding Agent that Runs in Your Terminal},
  year   = {2026},
  url    = {https://github.com/openai/codex}
}

@misc{anthropic2026claudecode,
  author = {{Anthropic}},
  title  = {{Claude Code}: An Agentic Coding Tool},
  year   = {2026},
  url    = {https://github.com/anthropics/claude-code}
}

@misc{qwen2026qwencode,
  author = {{Qwen Team}},
  title  = {{Qwen Code}: An Open-Source {AI} Agent for the Terminal},
  year   = {2026},
  url    = {https://github.com/QwenLM/qwen-code}
}

@misc{webalmanac2022css,
  author = {{HTTP Archive}},
  title = {The 2022 Web Almanac: {CSS} Chapter},
  year = {2022},
  howpublished = {HTTP Archive Web Almanac},
  url = {https://almanac.httparchive.org/en/2022/css}
}

@misc{webalmanac2025a11y,
  author       = {{HTTP Archive}},
  title        = {The 2025 Web Almanac: Accessibility Chapter},
  year         = {2025},
  howpublished = {HTTP Archive Web Almanac},
  url          = {https://almanac.httparchive.org/en/2025/accessibility}
}

@misc{stackoverflow2025survey,
  author       = {{Stack Overflow}},
  title        = {2025 Developer Survey: {AI}},
  year         = {2025},
  url          = {https://survey.stackoverflow.co/2025/ai}
}

@inproceedings{chang1993animation,
  author    = {Chang, Bay-Wei and Ungar, David},
  title     = {Animation: From Cartoons to the User Interface},
  booktitle = {Proceedings of the 6th Annual ACM Symposium on User Interface Software and Technology (UIST)},
  pages     = {45--55},
  year      = {1993},
  doi       = {10.1145/168642.168647}
}

@article{heer2007animated,
  author  = {Heer, Jeffrey and Robertson, George G.},
  title   = {Animated Transitions in Statistical Data Graphics},
  journal = {IEEE Transactions on Visualization and Computer Graphics},
  volume  = {13},
  number  = {6},
  pages   = {1240--1247},
  year    = {2007},
  doi     = {10.1109/TVCG.2007.70539}
}

@misc{puppeteer,
  author       = {{Puppeteer Contributors}},
  title        = {{Puppeteer}: JavaScript API for Chrome and Firefox},
  howpublished = {\url{https://github.com/puppeteer/puppeteer}},
  note         = {Accessed: 2026-10-21}
}

@misc{awwwards,
  author       = {{Awwwards}},
  title        = {Animated Websites},
  howpublished = {\url{https://www.awwwards.com/websites/animations/}},
  note         = {Accessed: 2026-10-21}
}

@techreport{gpt2,
  title       = {Language Models are Unsupervised Multitask Learners},
  author      = {Radford, Alec and Wu, Jeffrey and Child, Rewon and
                 Luan, David and Amodei, Dario and Sutskever, Ilya},
  institution = {OpenAI},
  year        = {2019},
  url         = {https://cdn.openai.com/better-language-models/language_models_are_unsupervised_multitask_learners.pdf}
}

@misc{easyocr,
  author       = {{JaidedAI}},
  title        = {EasyOCR},
  howpublished = {\url{https://github.com/JaidedAI/EasyOCR}},
  year         = {2020}
}

@inproceedings{gui2025UICopilot,
author = {Gui, Yi and Wan, Yao and Li, Zhen and Zhang, Zhongyi and Chen, Dongping and Zhang, Hongyu and Su, Yi and Chen, Bohua and Zhou, Xing and Jiang, Wenbin and Zhang, Xiangliang},
title = {UICopilot: Automating UI Synthesis via Hierarchical Code Generation from Webpage Designs},
year = {2025},
isbn = {9798400712746},
publisher = {Association for Computing Machinery},
address = {New York, NY, USA},
url = {https://doi.org/10.1145/3696410.3714891},
doi = {10.1145/3696410.3714891},
pages = {1846–1855},
numpages = {10},
location = {Sydney NSW, Australia},
series = {WWW '25}
}

@article{wan2024automatically,
author = {Wan, Yuxuan and Wang, Chaozheng and Dong, Yi and Wang, Wenxuan and Li, Shuqing and Huo, Yintong and Lyu, Michael},
title = {Divide-and-Conquer: Generating UI Code from Screenshots},
year = {2025},
issue_date = {July 2025},
publisher = {Association for Computing Machinery},
address = {New York, NY, USA},
volume = {2},
number = {FSE},
url = {https://doi.org/10.1145/3729364},
doi = {10.1145/3729364},
journal = {Proc. ACM Softw. Eng.},
month = jun,
articleno = {FSE094},
numpages = {24}
}

@article{xiao2026interwebcoder,
  title={InterWebCoder: Learning Interactive Webpage Generation from Executable Tests via Reinforcement Learning},
  author={Xiao, Jingyu and Dai, Dasen and Wan, Yuxuan and Huang, Jen-tse and Huo, Yintong and Zhang, Hongyu and Lyu, Michael R},
  journal={arXiv preprint arXiv:2610.28610},
  year={2026}
}

@article{zhou2024bridging,
author = {Zhou, Ting and Zhao, Yanjie and Hou, Xinyi and Sun, Xiaoyu and Chen, Kai and Wang, Haoyu},
title = {DeclarUI: Bridging Design and Development with Automated Declarative UI Code Generation},
year = {2025},
issue_date = {July 2025},
publisher = {Association for Computing Machinery},
address = {New York, NY, USA},
volume = {2},
number = {FSE},
url = {https://doi.org/10.1145/3715726},
doi = {10.1145/3715726},
journal = {Proc. ACM Softw. Eng.},
month = jun,
articleno = {FSE011},
numpages = {23}
}

@article{wu2024mllm,
author = {Wu, Fan and Gao, Cuiyun and Li, Shuqing and Wen, Xin-Cheng and Liao, Qing},
title = {MLLM-Based UI2Code Automation Guided by UI Layout Information},
year = {2025},
issue_date = {July 2025},
publisher = {Association for Computing Machinery},
address = {New York, NY, USA},
volume = {2},
number = {ISSTA},
url = {https://doi.org/10.1145/3728925},
doi = {10.1145/3728925},
journal = {Proc. ACM Softw. Eng.},
month = jun,
articleno = {ISSTA050},
numpages = {23}
}

@article{xu2026video2code,
  title={Video2Code: Generating Interactive Webpages from UI Videos via Action-Aware Revisit},
  author={Xu, Mingde and Yang, Zhen and Wang, Yan and Wang, Yu and Liu, Xijun and Dou, Zijun and Hong, Wenyi and Gu, Xiaotao and Xu, Bin and Tang, Jie},
  journal={arXiv preprint arXiv:2606.20711},
  year={2026}
}

@article{xiao2026efficientuicoder,
  title={EfficientUICoder: A Bidirectional Token Compression Framework for Efficient MLLM-Based UI Code Generation},
  author={Xiao, Jingyu and Zhang, Zhongyi and Wan, Yuxuan and Huo, Yintong and Liu, Yang and Lyu, Michael R},
  journal={Proceedings of the ACM on Software Engineering},
  volume={3},
  number={FSE},
  pages={2396--2418},
  year={2026},
  publisher={ACM, New York, NY, USA}
}









\clearpage
\appendix
\section{Data Copyright and Usage}
\label{app:licensing}

The webpages in AnimWeb-Bench were collected from Awwwards~\cite{awwwards}, and all webpages, recordings, and assets remain the property of their respective owners.
To respect their ownership, we retain the source URL of each instance to credit the original webpage, and release AnimWeb-Bench solely for non-commercial research purposes.
We will also promptly remove any content found to be used inappropriately.

\section{Dataset Statistics}

Table~\ref{tab:dataset-stats} reports the statistics of AnimWeb-Bench. Each video lasts 5.91\,s and contains 2.83 animation segments and 35.4 tracklets on average, and each sample comes with 46.1 asset files on average.

Following Design2Code~\cite{si2025design2code}, we measure code complexity with four metrics: Length is the token count of the HTML code under the GPT-2 tokenizer~\cite{gpt2}, Tags is the number of HTML tags, DOM Depth is the maximum depth of the DOM tree, and Unique Tags is the number of distinct tags. As shown in Table~\ref{tab:complexity-compare}, an Animation2Code~\cite{ji2026animation2codeevaluatingtemporalvisual} sample contains only 71 tags on average, with a DOM depth of 5 and 7 distinct tags, since each sample is a standalone animation snippet rather than a complete webpage. In contrast, AnimWeb-Bench pages contain 1,196 tags, a DOM depth of 18, and 33 distinct tags on average, exceeding all the compared webpage benchmarks.

\section{Implementation Details}
\label{app:implementation}

This section specifies the terms and thresholds used in the evaluation metrics of Section~3.3.

\paragraph{Keyframe alignment.}
On downsampled frames, $\mathrm{Vis}$ in Eq.~\eqref{eq:kf} is computed as
\begin{equation*}
\mathrm{Vis} = 0.15\,s_{\mathrm{rgb}} + 0.10\,s_{\mathrm{gray}} + 0.10\,s_{\mathrm{hist}} + 0.10\,s_{\mathrm{edge}} + 0.10\,s_{\mathrm{cos}} + 0.45\,s_{\mathrm{layout}},
\end{equation*}
where $s_{\mathrm{rgb}}$ and $s_{\mathrm{gray}}$ are one minus the mean absolute pixel differences in RGB and grayscale, $s_{\mathrm{hist}}$ is the RGB histogram intersection, $s_{\mathrm{edge}}$ and $s_{\mathrm{cos}}$ are the cosine similarities of Canny edge maps and of RGB images, and $s_{\mathrm{layout}}$ compares the connected regions of white, colored, and foreground masks. We set $\lambda = 0.85$. Keyframes are aligned by order-preserving dynamic programming, where a pair can be matched only if $s \geq 0.2$, and each unmatched reference keyframe incurs a penalty of 0.12.

\paragraph{Structure similarity.}
All boxes are normalized by the frame width and height. For a matched pair $(b^r, b^g)$ in Eq.~\eqref{eq:structure-pair}, $s_{\mathrm{iou}}$ is the intersection over union of the two boxes; $s_{\mathrm{center}}$ is the center proximity computed as in Eq.~\eqref{eq:pos}, applied to the centers of $b^r$ and $b^g$; $s_{\mathrm{area}}$ is the ratio of the smaller box area to the larger one; and $s_{\mathrm{aspect}}$ is the ratio of the smaller aspect ratio to the larger one.

\paragraph{Trajectory construction and matching.}
In Algorithm~\ref{alg:trajectory}, $\mathrm{IoU}$ is the intersection over union between a detection $b$ and the latest box of trajectory $\tau$, $\mathrm{App}$ is the cosine similarity between the CLIP image features~\cite{radford2021learning} of the two box regions, and $\mathrm{Ctr}$ is the center proximity computed as in Eq.~\eqref{eq:pos}. We set $\theta_a = 0.35$ and $\theta_t = 0.82$.

\begin{table}[!t]
\centering
\caption{Video and asset statistics of AnimWeb-Bench.}
\label{tab:dataset-stats}
\small
\setlength{\tabcolsep}{10pt}
\begin{tabular}{lrrr}
\toprule
\textbf{Metric} & \textbf{Min} & \textbf{Max} & \textbf{Mean $\pm$ Std} \\
\midrule
Duration (s) & 2.0 & 15.0 & 5.91\,{\scriptsize$\pm$2.47} \\
Segments & 1 & 11 & 2.83\,{\scriptsize$\pm$1.84} \\
Tracklets & 2 & 120 & 35.4\,{\scriptsize$\pm$24.1} \\
Asset files & 2 & 422 & 46.1\,{\scriptsize$\pm$53.6} \\
\bottomrule
\end{tabular}
\end{table}

\begin{table*}[!t]
\centering
\caption{Code complexity of AnimWeb-Bench and existing benchmarks (Mean $\pm$ Std).}
\label{tab:complexity-compare}
\small
\setlength{\tabcolsep}{8pt}
\begin{tabular}{lccccc}
\toprule
\textbf{Metric} & \textbf{Design2Code} & \textbf{WebCode2M} & \textbf{MRWeb} & \textbf{Animation2Code} & \textbf{AnimWeb-Bench} \\
\midrule
Length (tokens) & 31,216\,{\scriptsize$\pm$23,902} & 5,366\,{\scriptsize$\pm$2,393} & 113,724\,{\scriptsize$\pm$139,761} & 5,427\,{\scriptsize$\pm$13,678} & \textbf{239,303}\,{\scriptsize$\pm$487,624} \\
Tags & 158\,{\scriptsize$\pm$100} & 184\,{\scriptsize$\pm$77} & 543\,{\scriptsize$\pm$768} & 71\,{\scriptsize$\pm$247} & \textbf{1,196}\,{\scriptsize$\pm$894} \\
DOM Depth & 13\,{\scriptsize$\pm$5} & 15\,{\scriptsize$\pm$5} & 15\,{\scriptsize$\pm$6} & 5\,{\scriptsize$\pm$3} & \textbf{18}\,{\scriptsize$\pm$4} \\
Unique Tags & 22\,{\scriptsize$\pm$6} & 24\,{\scriptsize$\pm$6} & 22\,{\scriptsize$\pm$8} & 7\,{\scriptsize$\pm$2} & \textbf{33}\,{\scriptsize$\pm$7} \\
\bottomrule
\end{tabular}
\end{table*}

\section{Correlation Between Automatic Metrics and Human Judgments}
\label{app:human-correlation}

\begin{table}[!t]
\centering
\caption{Spearman's rank correlation $\rho$ between automatic metrics and human preferences over 400 pairwise comparisons against the baseline.}
\label{tab:human-correlation}
\small
\setlength{\tabcolsep}{8pt}
\renewcommand{\arraystretch}{1.1}
\begin{tabular}{@{}llc@{}}
\toprule
\textbf{Dimension} & \textbf{Metric} & \textbf{Spearman's $\rho$} \\
\midrule
\multirow{4}{*}{Static Appearance Fidelity}
 & Structure & 0.7365 \\
 & Text      & 0.7113 \\
 & Color     & 0.6951 \\
 & CLIP      & 0.7172 \\
\midrule
\multirow{4}{*}{Animation Fidelity}
 & Position  & 0.7138 \\
 & Size      & 0.7350 \\
 & Velocity  & 0.6667 \\
 & CLIP      & 0.7160 \\
\bottomrule
\end{tabular}
\end{table}

We quantify the agreement between our automatic metrics and human judgments using Spearman's rank correlation coefficient $\rho$.
The human evaluation in Section~\ref{sec:rq4} yields 400 pairwise comparisons (8 models $\times$ 50 webpages), each between a model and the baseline.
For each comparison, we encode the majority-vote human preference as $1$ if the model is preferred, $0$ for a tie, and $-1$ if the baseline is preferred, and compute the difference between the model's and the baseline's automatic scores on that webpage.
Each metric is correlated with the human preferences in its own dimension, i.e., static metrics with static appearance judgments and animation metrics with animation judgments.
Ties are retained and assigned average ranks.
As shown in Table~\ref{tab:human-correlation}, all eight metrics correlate positively with human preferences ($\rho = 0.667$--$0.737$), supporting their alignment with human judgments.

\section{Failure Case Studies}
\label{app:case_study}


\begin{figure*}[!t]
  \centering
  \includegraphics[width=\textwidth]{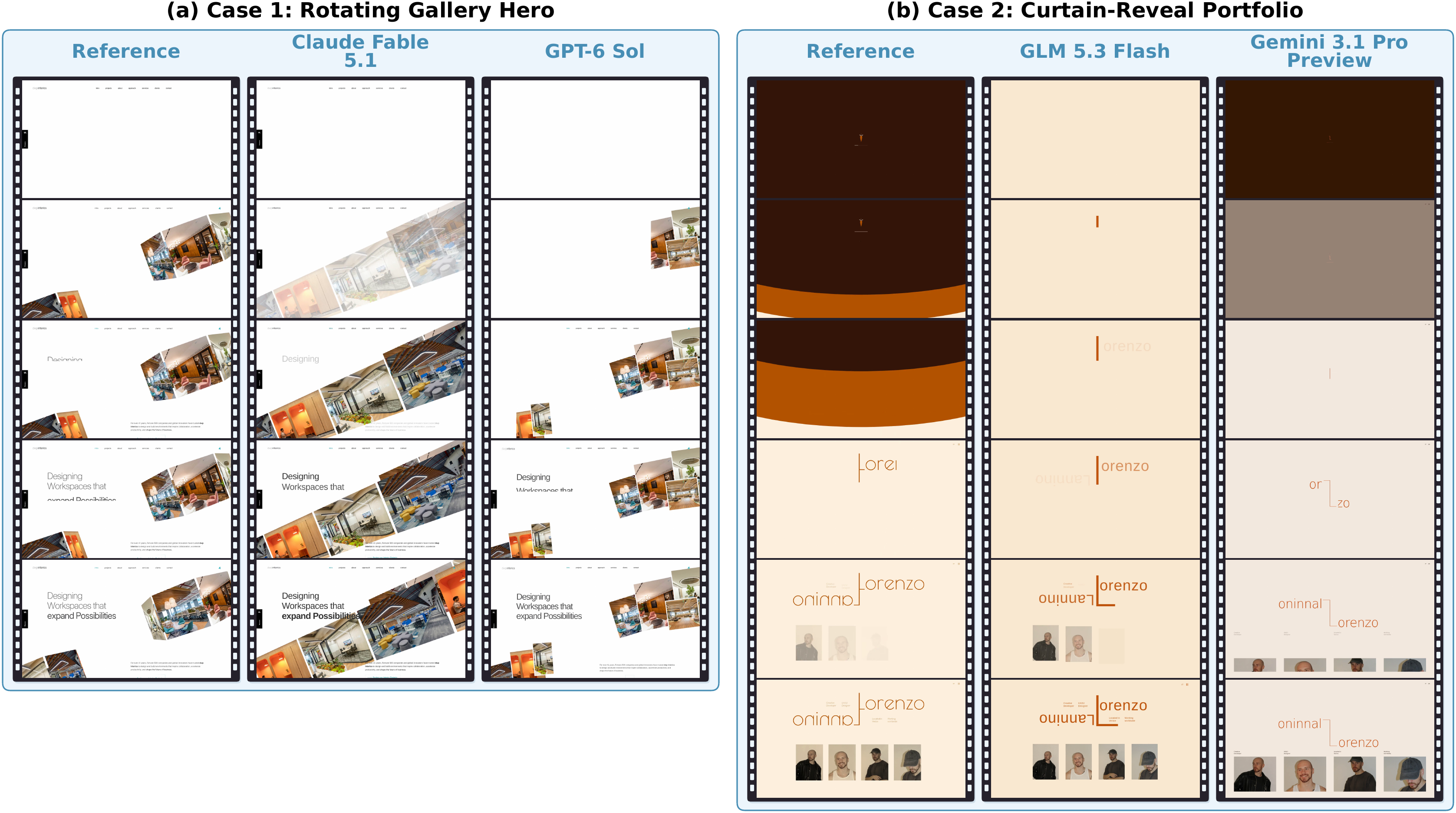}
  \caption{Failure cases on two webpages. Frames are ordered from top to bottom in time.}
  \label{fig:case-study}
\end{figure*}

\subsection{Case 1: Rotating Gallery Hero}

In the reference (Figure~\ref{fig:case-study}(a)), the navigation bar is present from the start. Two photo clusters at the top-right and bottom-left corners then rotate in depth, the paragraph at the bottom right appears, and the headline is finally revealed line by line through a mask.

\textbf{Claude Fable 5.1} merges the two clusters into a single diagonal strip spanning the page, which lowers Structure. The strip fades in and drifts toward the bottom left instead of rotating in place, so its tiles deviate in Position, Size, and animation CLIP, and move in a different direction and at a different speed, lowering Velocity. The headline fades in rather than emerging from a mask, so its box appears at full size instead of growing as it is unveiled, further lowering Size.

\textbf{GPT-6 Sol} reproduces the masked headline but not the rotation: its photos become small, static fans of cards, lowering Size and Velocity. Its timeline is also reordered: the photos appear before the navigation bar, and the paragraph only after the headline. Components shown in the reference are therefore missing from the generated page for many time steps, which lowers all four animation metrics.


\subsection{Case 2: Curtain-Reveal Portfolio}

In the reference (Figure~\ref{fig:case-study}(b)), a loading icon animates on a dark screen. A curtain-like background transition follows, in which brown and orange layers sweep upward. The name is then revealed letter by letter from left to right, and the four photos fade in place one after another.

\textbf{GLM 5.3 Flash} omits the loading screen and the transition entirely. The first reference keyframe and segment therefore have no counterpart, lowering all static and animation metrics. The name fades in as whole words instead of being revealed letter by letter, lowering all four animation metrics. The photos rise from below instead of fading in place, lowering Position and Velocity.

\textbf{Gemini 3.1 Pro Preview} keeps the loading screen but not its animated icon. Its transition fades the background color instead of sweeping the layers upward, so the moving layers find no counterpart, lowering all four animation metrics. The two words of the name swap their vertical order, and the text is misplaced, lowering Structure and static CLIP. The four photos slide up from below all at once instead of fading in one after another, lowering all four animation metrics. They also settle enlarged at the wrong positions, further lowering Structure and static CLIP.

\section{Coding Agent Execution Trace}
\label{app:trace}

\begin{figure*}[t]
  \centering
  \includegraphics[width=\textwidth]{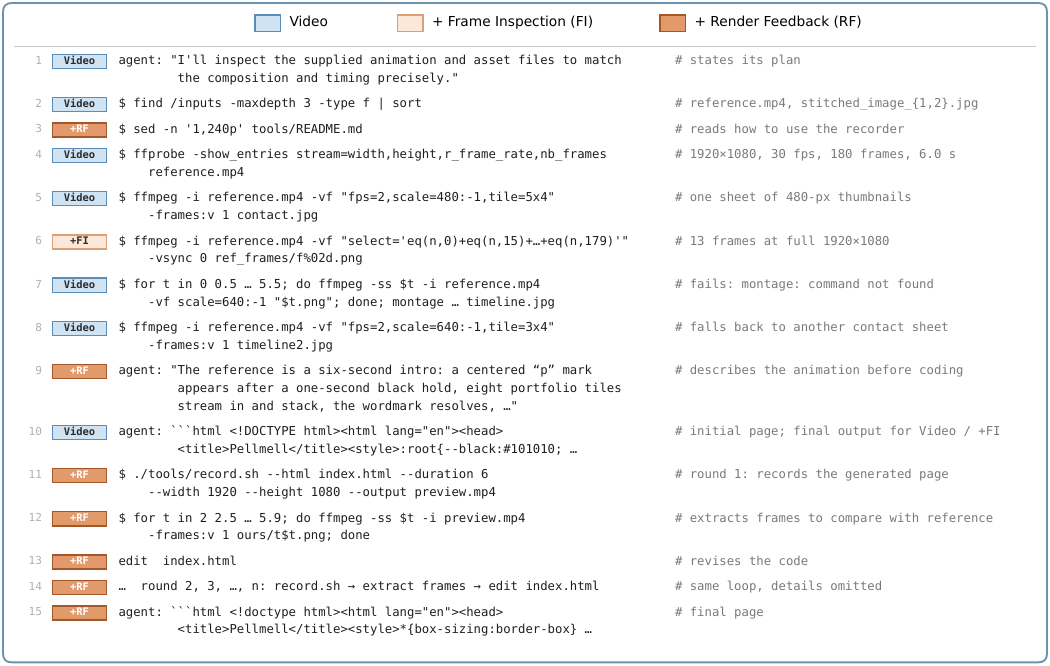}
  \caption{Simplified Codex execution log on one webpage of AnimWeb-Bench across three settings. Lines tagged +FI or +RF are steps added by that setting.}
  \label{fig:codex-trace}
\end{figure*}

Figure~\ref{fig:codex-trace} shows a simplified Codex execution log on one webpage across the three settings.
The settings are cumulative: FI adds frame inspection to video input, and RF further adds render feedback.
With video input alone, the agent perceives the animation only through contact sheets of downscaled frames (lines 5--8) and returns its first page without ever viewing it (line 10).
Frame Inspection supplies the same moments at full resolution (line 6), exposing details lost in the thumbnails, but the generated page still goes unseen.
Render Feedback closes this loop: the agent records its own page (line 11), extracts frames to compare with the reference (line 12), and revises the code (line 13), repeating this cycle before returning the final page (lines 14--15).
Only under Render Feedback does the agent observe the output it generates, allowing errors in timing and motion to be corrected.

\end{document}